\documentclass[3p]{elsarticle}
\usepackage{float}
\usepackage{lineno,hyperref,color, soul}
\usepackage{amsmath}
\usepackage{amssymb}
\usepackage{natbib}
\usepackage{mathtools}
\usepackage{pdflscape}
\usepackage[font=small]{caption}
\usepackage[utf8]{inputenc}
\usepackage[T1]{fontenc}
\usepackage{multirow}

\usepackage{tikz}
\usetikzlibrary{patterns}
\usetikzlibrary{arrows}
\modulolinenumbers[5]

\newcommand{\delete}[1]{} 

\newcommand*\patchAmsMathEnvironmentForLineno[1]{%
  \expandafter\let\csname old#1\expandafter\endcsname\csname #1\endcsname
  \expandafter\let\csname oldend#1\expandafter\endcsname\csname end#1\endcsname
  \renewenvironment{#1}%
  {\linenomath\csname old#1\endcsname}%
  {\csname oldend#1\endcsname\endlinenomath}}%
\newcommand*\patchBothAmsMathEnvironmentsForLineno[1]{%
  \patchAmsMathEnvironmentForLineno{#1}%
  \patchAmsMathEnvironmentForLineno{#1*}}%
\AtBeginDocument{%
  \patchBothAmsMathEnvironmentsForLineno{equation}%
  \patchBothAmsMathEnvironmentsForLineno{align}%
  \patchBothAmsMathEnvironmentsForLineno{flalign}%
  \patchBothAmsMathEnvironmentsForLineno{alignat}%
  \patchBothAmsMathEnvironmentsForLineno{gather}%
  \patchBothAmsMathEnvironmentsForLineno{multline}%
}

\journal{Spatial Statistics}

\biboptions{authoryear}

\begin{document}
\begin{frontmatter}



\title{Identification of Dominant Features in Spatial Data}

\author[add1]{Roman Flury}\ead{roman.flury@math.uzh.ch}
\author[add2]{Florian Gerber}
\author[add3]{Bernhard Schmid}
\author[add1,add4]{Reinhard Furrer}\ead{reinhard.furrer@math.uzh.ch}

\address[add1]{Department of Mathematics, University of Zurich, Zurich, Switzerland}
\address[add2]{Department of Applied Mathematics and Statistics, Colorado School of Mines, Golden, USA}
\address[add3]{Department of Geography, University of Zurich, Zurich, Switzerland}
\address[add4]{Department of Computational Science, University of Zurich, Zurich, Switzerland}

\begin{abstract}
  Dominant features of spatial data are connected structures or patterns that emerge from location-based variation and manifest at specific scales or resolutions.
  To identify dominant features, we propose a sequential application of multiresolution decomposition and variogram function estimation.
  Multiresolution decomposition separates data into additive components, and in this way enables the recognition of their dominant features.
  A dedicated multiresolution decomposition method is developed for arbitrary gridded spatial data, where the underlying model includes a precision and spatial-weight matrix to capture spatial correlation.
  The data are separated into their components by smoothing on different scales, such that larger scales have longer spatial correlation ranges.
  Moreover, our model can handle missing values, which is often useful in applications.
  Variogram function estimation can be used to describe properties in spatial data.
  Such functions are therefore estimated for each component to determine its effective range, which assesses the width-extent of the dominant feature.
  Finally, Bayesian analysis enables the inference of identified dominant features and to judge whether these are credibly different.
  The efficient implementation of the method relies mainly on a sparse-matrix data structure and algorithms.
  By applying the method to simulated data we demonstrate its applicability and theoretical soundness.
  In disciplines that use spatial data, this method can lead to new insights, as we exemplify by identifying the dominant features in a forest dataset.
  In that application, the width-extents of the dominant features have an ecological interpretation, namely the species interaction range, and their estimates support the derivation of ecosystem properties such as biodiversity indices.
\end{abstract}


\end{frontmatter}

\section{Introduction}\label{sec:introduction}
  In various scientific disciplines, including ecology, epidemiology, climate science, hydrology, and fluid dynamics, relevant questions are often associated with specific scale-dependent features of the data.
  In spatial data, features are connected structures or patterns that emerge from location-based variation.
  Features occur for various reasons, e.g. local differences in topographic features that affect tree heights in a forest, or globally varying climatic conditions that create temperature patterns.
  These features usually depend on the data's resolution or geographical extent, denoted as scale, respectively, as scale-dependence.
  Typically, one expects to recognize different features on noise or micro, local, regional or global scales~\citep{Wikle2005, Wu2013}.
  For a complete statistical analysis, it is essential to identify these scale-dependent features and thus to detect all relevant scales.
  Assuming that each scale is characterized by dominant features, the assessment of properties of these dominant features, such as the \emph{feature width-extent}, is of equal importance.
  These widths help understand the scale-dependent features themselves and their relationship to the underlying mechanisms~\citep{Delcourt88, Skoien2003, Pasanen2018}.
  We understand the whole procedure of detecting scales, recognizing the dominant scale-dependent features, and assessing their extents as \textit{feature identification}.

  To find relevant scales in spatial data, in the past, signal-decomposition and image-processing methods have been applied.
  Decomposition approaches were first proposed in the computer-vision literature \citep{Witkin83, Lindeberg94} and then adapted in statistical scale-space analysis, in which observations interpreted as realizations of a random variable are considered at several scales.
  The scales are then obtained through smoothing using a moving average~\citep{Chaudhuri99}.
  These ideas were extended by~\citet{Erasto2005} and are now known as \emph{Bayesian significant zero crossings of derivatives} and further developed to the \emph{scale-space multiresolution analysis} by~\cite{Holmstrom2011}, for a general overview see also~\cite{Holmstrom2017}.
  The main goal of the scale-space multiresolution analysis is to recognize scale-dependent features from time-series data and images, as was successfully demonstrated in several applied studies~\citep{Lehmann2017, Aakala2018, Kulha2019}.

  Methods to assess \emph{feature width-extents} directly from the data are used in spectral analysis, introduced by~\citet{Ford84} and applied, for instance, in~\citet{Stoll94}.
  However, \citet{Pasanen2018} were the first to describe an approach to determine the actual size of scale-dependent features.
  They showed that the maximum of the so-called scale-derivative norm~\citep{Pasanen2013} of a recognized component can be used to estimate its \emph{characteristic feature size}.

  In spatial statistics, data properties are often described with variogram or covariance functions and their respective parameters.
  Depending on the dimension of the data, the approximation of their parameters becomes computationally intensive.
  Therefore, \emph{spatial multiresolution models} have been developed to reduce the resolution in the data while preserving their information content.
  These models often use a linear combination of basis functions to model spatially dependent data.
  The coefficients are then usually estimated to weight the respective basis functions~\citep{Cressie2008, Katzfuss2012, Nychka2015}, or chosen to approximate a given covariance function optimally~\citep{Katzfuss2017}.

  In this article, we first develop a multiresolution decomposition for spatial data on the basis of the statistical scale-space method of~\citet{Holmstrom2011} (Section~\ref{sec:mresd}).
  This scale-space analysis and its implementation are only applicable to complete and regularly gridded data.
  However, spatial data typically contains dependencies along any direction and their intensity is usually strongly linked to the distance between locations and becomes weaker as the distance increases~\citep{Cressie93}.
  Depending on the spatial setting, it is also possible that there is no dependency between specific locations.
  Therefore, we propose a decomposition method that enables accurate modeling of the dependency and neighboring relations between locations.
  The newly developed method is applicable for arbitrary gridded spatial data, where the resolution of the grid points defines the aggregation of the represented area.
  With this method, it is possible to exclude specific grid locations that are not relevant for the analysis or without assigned value.
  Accounting for the latter case, we introduce a flexible procedure to resample missing values in the data.

  In addition, we propose assessing the width-extents of the dominant scale-dependent features (Section~\ref{subsec:featsize}).
  To do so, we estimate the empirical variogram~\citep{Matheron62, Cressie93} and optimize a Mat\'ern variogram function for its parameters.
  We assess the width-extent of the individual scale-dependent features with its respective effective range parameter~\citep{Pebesma2004, Nychka2020}.
  Importantly, unlike in \emph{spatial multiresolution models}, we do not reduce the dimensionality in the data, while preserving their information content to approximate a covariance function efficiently.
  But by approximating a variogram function based on each component separately, we identify the width-extents of the dominant scale-dependent feature in largest-possible isolation.

  We first apply the new method to simulated data (Section~\ref{sec:illustration}).
  In this setting, we can control the parameters of the additive spatial data and demonstrate the consistency of the results.
  Second, we demonstrate the usefulness of the method by an application (Section~\ref{sec:application}) in which we explain how the identification of dominant features can be used to find different area sizes in which communities of species interact with each other~\citep{Greig79}.
  The  diameter of such a relevant community of species is known as the interaction range.
  It can be used to determine different radii for multidimensional biodiversity indices based on moving-window approaches.
  In this application, we use scientifically relevant data, which were obtained by remote sensing from the forest on the hillslope of mountain Laegeren in Switzerland.
  For this area, multidimensional functional biodiversity indices were calculated in previous studies and evaluated at different radii determined by experts.

\section{Feature identification}\label{sec:mresd}
\subsection{Spatial data resampling based on a Bayesian hierarchical model}\label{subsec:resamp}
  As in the scale-space analysis method, we assume that the observed dataset $\boldsymbol{y}$ is a composition of the true underlying data $\boldsymbol{x}$ and additional white noise $\boldsymbol{\varepsilon}$ with constant variance.
  To separate observational noise from the unobserved data, $\boldsymbol{y}$ is modeled and resampled using a Bayesian model~\citep{Gelfand2012}.
  For the inferred features, credibility maps can then be derived to distinguish noise from these features~\citep{Erasto2005}.
  In the scale-space analysis, the Bayesian model is constructed such that its posterior distribution is of closed form and that it is possible to calculate the inverse of the precision matrix by fast Fourier transformation~\citep{Strang99, Reuter2009}.
  However, these computational gains restrict this method to complete and regularly gridded data with a fixed precision matrix.
  To remove these restrictions on the data and enable arbitrary precision matrices, we need to replace the efficient fast Fourier transformation implementation.
  As precision matrices are typically sparse, we can rely on its economic data structure and efficient algorithms.
  For the statistical software \textsc{R} those are implemented, for example, in the \textbf{spam} package~\citep{Furrer2010}.
  Having the possibility to represent the precision matrix of extensive spatial data, the calculation of its inverse --- the covariance matrix --- remains computationally expensive (see Section~\ref{subsec:compdetails} for more computational details).
  For this reason, we choose a normal-gamma model to resample spatial data~\citep{Rue2005}.
  With this model, it is possible to sample from the canonical representation of a multivariate normal distribution, where the inverse of the precision matrix does not need to be calculated explicitly.

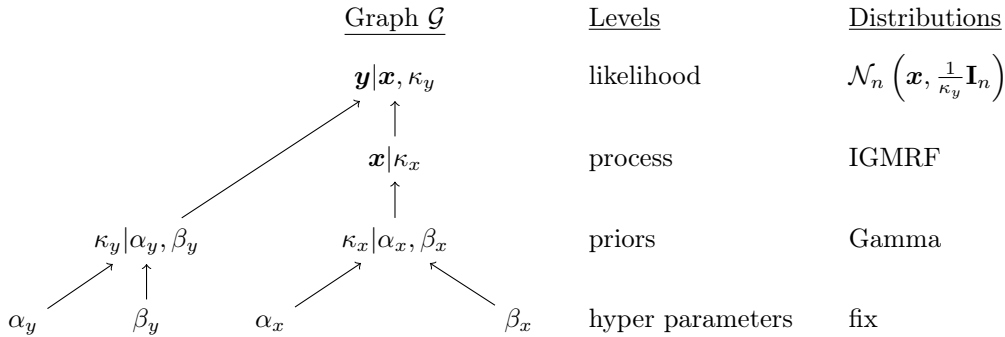
\begin{figure}
  \begin{center}
    \begin{tikzpicture}[scale=2]
      \tikzstyle{ann} = [draw=none,fill=none,right]
      \matrix[column sep=0.5cm,
      row 1/.style={yshift =.2cm},
      row 2/.style={yshift =.7cm},
      row 3/.style={yshift =.7cm},
      row 4/.style={yshift =.7cm},
      column 1/.style={anchor=base},
      column 2/.style={anchor=base},
      column 3/.style={anchor=base},
      column 4/.style={anchor=base},
      column 5/.style={anchor=base},
      column 6/.style={anchor=base west},
      column 7/.style={anchor=base west}] {
      &&&\node (titleG) {\underline{Graph $\mathcal{G}$}};&&
         \node (titleN) {\underline{Levels}};&
         \node (titleD) {\underline{Distributions}};\\
      &&&\node (1)  {$\boldsymbol{y}|\boldsymbol{x}, \kappa_y$};&&
         \node (9)  {likelihood};&
         \node (10) {$\mathcal{N}_n \left( \boldsymbol{x}, \frac{1}{\kappa_y}\mathbf{I}_n \right)$};\\
      &&&\node (2)  {$\boldsymbol{x}|\kappa_x$};&&
         \node (11) {process};&
         \node (12) {IGMRF};\\
        &\node (4)  {$\kappa_y|\alpha_y,\beta_y$};&&
         \node (3)  {$\kappa_x|\alpha_x,\beta_x$};&&
         \node (13) {priors};&
         \node (14) {Gamma};\\
         \node (5)  {$\alpha_y$};&
         \node (6)  {$\beta_y$};&
         \node (7)  {$\alpha_x$};&&
         \node (8)  {$\beta_x$};&
         \node (15) {hyper parameters};&
         \node (16) {fix};\\
      };
      \draw [<-] (1) to (2);
      \draw [<-] (1) to (4);
      \draw [<-] (2) to (3);
      \draw [<-] (3) to (7);
      \draw [<-] (3) to (8);
      \draw [<-] (4) to (5);
      \draw [<-] (4) to (6);
    \end{tikzpicture}
    \caption{A graphical-model representation of the Bayesian hierarchical normal-gamma model to resample the observed spatial dataset $\boldsymbol{y}$.}
    \label{fig:model}
  \end{center}
\end{figure}

  The normal-gamma model includes a multivariate normal likelihood function for the observed dataset $\boldsymbol{y}$ at conditionally independent locations, with the true mean $\boldsymbol{x}$ and precision parameter $\kappa_{\boldsymbol{y}}$.
  It is proportional to
\begin{equation}
  \pi(\boldsymbol{y}|\boldsymbol{x}, \kappa_y) \propto \kappa_y^{n/2} \exp\left( - \frac{\kappa_y}{2} \left(\boldsymbol{y} - \boldsymbol{x} \right)^\top \left( \boldsymbol{y} - \boldsymbol{x} \right) \right)\nonumber,
\end{equation}
  where $n$ denotes the total number of locations in the spatial data.
  To model the spatial dependencies of $\boldsymbol{x}$, we use an intrinsic Gaussian Markov Random Field (IGMRF) with zero mean, a precision parameter $\kappa_x$, and a spatial-weight matrix $\boldsymbol{Q}$, which represents the dependencies between locations.
\begin{equation}\label{eq:igmrf}
  \pi(\boldsymbol{x} | \kappa_x) \propto \kappa_x^{(n-2)/2} \exp\left( -\frac{\kappa_x}{2} \boldsymbol{x}^\top\boldsymbol{Q}\boldsymbol{x} \right).
\end{equation}
  For the unknown precision parameters of the normal likelihood and the spatial process, we choose independent Gamma distributions, the respective conjugate prior distributions.
  These Gamma priors have strictly positive shape and rate hyperparameters $\alpha_x, \alpha_y, \beta_x$, and $\beta_y$.
  The final Bayesian hierarchical model is summarized in Fig.~\ref{fig:model}.
  The resulting conditional distribution $\pi\left( \boldsymbol{x}| \boldsymbol{y}, \kappa_y, \kappa_x \right)$ for the true underlying data $\boldsymbol{x}$, is a multivariate normal, where the conditional distributions for the unknown parameters of the normal-gamma model can be identified as Gamma distributions.
\begin{align}
  \boldsymbol{x}|\boldsymbol{y}, \kappa_y, \kappa_x &\sim \mathcal{N}_{\mathcal{C}, n} \left( \kappa_y\boldsymbol{y}, \kappa_x\boldsymbol{Q} + \kappa_y \mathbf{I}_n \right),\label{eq:normalgammacanonical}\\
  \kappa_x|\boldsymbol{x}, \boldsymbol{y} &\sim \text{Gamma} \left( \alpha_x + \frac{n - 2}{2}, \beta_x + \frac{1}{2}\boldsymbol{x}^\top\boldsymbol{Q}\boldsymbol{x} \right),\label{eq:gammaQ}\\
  \kappa_y|\boldsymbol{x}, \boldsymbol{y} &\sim \text{Gamma} \left( \alpha_y + \frac{n}{2}, \beta_y + \left(\boldsymbol{y} - \boldsymbol{x} \right)^\top \left( \boldsymbol{y} - \boldsymbol{x} \right) \right).\label{eq:gammaQ2}
\end{align}
  In equation~\eqref{eq:normalgammacanonical}, $\mathbf{I}_{n}$ denotes the identity matrix with dimension $n$ and $\mathcal{N}_{\mathcal{C}, n}$ denotes the canonical parametrization of a multivariate normal distribution of dimension $n$.
  These three distributions can be identified as the full conditional distributions of the Bayesian hierarchical model.
  As these are of closed form, we can use an efficient Gibbs sampling approach to resample the spatial data~\citep{Gerber2015}.

  To construct a spatial-weight matrix $\boldsymbol{Q}$ for spatial data on a lattice, as used in equations~\eqref{eq:igmrf},~\eqref{eq:normalgammacanonical} and~\eqref{eq:gammaQ}, we assume that the underlying spatial process is an IGMRF of first order, that is, dependent on first-order neighbor locations.
  If a higher-order IGMRF fits the data more closely, these derivations work similarly.
  An IGMRF of first order is an improper GMRF with precision matrix of rank $n - 1$, such that each row-sum of the precision matrix is equal to zero.
  Moreover, an IGMRF is defined with respect to a labelled graph such that $\boldsymbol{Q}_{ij} \ne 0$ if and only if the locations $i$ and $j$ share a common edge in this graph, i.e., are neighbors.
  For a complete mathematical definition see~\cite{Rue2005}

\subsubsection{Regular gridded data}\label{sec:reggrid}
  If the data are observed on a regular grid $\mathcal{I}_n$ with $n = n_1n_2$ nodes, we can define
\begin{equation}\label{eq:regularQ}
  \boldsymbol{Q} = \alpha_1 \boldsymbol{R}_{n_1} \otimes \mathbf{I}_{n_2} + \alpha_2 \boldsymbol{R}_{n_2} \otimes \mathbf{I}_{n_1}.
\end{equation}
  $\boldsymbol{R}_n$ is the structure matrix of a random walk of order one and dimension $n$:
\begin{equation}
  \boldsymbol{R}_n = \begin{pmatrix}
       1 & -1     &        &        &    \\
      -1 &  2     &     -1 &        &    \\
         & \ddots & \ddots & \ddots &    \\
         &        &     -1 &  2     & -1 \\
         &        &        & -1     &  1 \\
  \end{pmatrix}\nonumber
\end{equation}
  The $\alpha$'s are constraint to $\alpha_1 + \alpha_2 = 2$, such that possible anisotropic behavior along the two axes can be specified.
  In an application, either $\alpha_1$ or $\alpha_2$ can be treated as another unknown parameter and thus the degree of anisotropy can be estimated from the data.
  More details on estimating these anisotropy parameters are given in~\cite{Rue2005}.

  Assuming that some grid locations do not belong to the area of interest, we need to modify the spatial-weight matrix $\boldsymbol{Q}$ from equation~\eqref{eq:regularQ}.
  Knowing the specific locations in the vectorized spatial data $\boldsymbol{x}$, we define $\boldsymbol{H}$ as a diagonal matrix with diagonal entries equal to one if the corresponding $(i, j)$ node of the lattice is to be considered in the analysis, and zero otherwise.
  The spatial-weight matrix is then calculated as $\boldsymbol{H}^\top\boldsymbol{Q}\boldsymbol{H}$, where $\boldsymbol{Q}$ is defined in equation~\eqref{eq:regularQ}.
  Furthermore, the diagonal entries of this matrix $\boldsymbol{H}^\top\boldsymbol{Q}\boldsymbol{H}$ have to be adjusted such that row sums are equal to zero.
  This approach can be used to ignore vast or small areas as well as to account for irregular boundaries of spatial data.
  It is a particular case of spatial data distributed on an irregular lattice.

\subsubsection{Irregular gridded data}\label{sec:irreggrid}
  For an arbitrary irregular lattice, the spatial-weight matrix can be constructed according to adjacency relations of the graph for the IGMRF.
  An adjacency relation is fulfilled if two nodes from this graph share an edge.
  Such relations can be represented with a square matrix, where its elements indicate whether pairs of vertices are adjacent or not~\citep{Seidel1968}.
  With this approach, it is possible to model the precision or covariance between locations based on their neighboring structure,~i.e., consider first or higher-order neighbors.
  Hence, not on the actual Euclidean distance between the locations but only the neighboring structure are relevant for the covariance.

\subsection{Resampling spatial data with missing values}
  We now assume that the observed dataset $\boldsymbol{y}$ contain missing values, which implies that the respective locations have no well-defined values allocated.
  Common approaches to deal with missing values are imputation methods, which use for instance the average value of the respective neighbors of a location to replace a missing value.
  However, for the final credibility analysis, the corresponding locations of the underlying data should manifest more uncertainty and lower credibility.
  To introduce higher uncertainty to locations with missing values, we propose to resample at locations with missing values in the spatial data resampling step, as described in Section~\ref{subsec:resamp}.
  We formalize this scenario by letting the observed data $\boldsymbol{y}$ contain $k$ arbitrarily missing values, where $k \in \mathbb{N}$, $0 < k < n$ and $n$ is the number of locations on the grid.
  We further refer here to $\boldsymbol{H}$ as the matrix operator from $\mathbb{R}^n \to \mathbb{R}^m$, such that $m = n - k$ is the number of not-missing or observed values.
  The observed data is $\boldsymbol{y} \in \mathbb{R}^m$ and $\boldsymbol{H}^\top \in \mathbb{R}^{n\times m}$ is therefore the identity mapping for the $m$ non-missing values of the observed dataset $\boldsymbol{y}$.
  With the help of $\boldsymbol{H}$ we can transform the likelihood function of the model to $\boldsymbol{y}|\boldsymbol{H}\boldsymbol{x}, \kappa_y \sim \mathcal{N}_m \left( \boldsymbol{H}\boldsymbol{x}, \frac{1}{\kappa_y} \mathbf{I}_m \right)$.
  The original composition of the observed data $\boldsymbol{y} = \boldsymbol{x} + \boldsymbol{\varepsilon}$ in Section~\ref{subsec:resamp} translates then to $\boldsymbol{y} = \boldsymbol{H}\boldsymbol{x} + \boldsymbol{\varepsilon}$, where only the non-missing observed data affect the posterior samples $\boldsymbol{x}$.
  The missing values are treated differently and are only sampled based on close neighbors' values and the prior influence.
  Hence, the further away a location with missing values is to locations with observed values, the larger the variance of the posterior sample for this location is.
  We adjust the hierarchical model in Fig.~\ref{fig:model} by replacing $\boldsymbol{x}$ with $\boldsymbol{Hx}$.
  The full conditional distributions, which define the Gibbs sampler, are adjusted to
\begin{align}
  \boldsymbol{x}|\boldsymbol{y},\kappa_y, \kappa_x &\sim \mathcal{N}_{\mathcal{C}, n} \left( \kappa_y(\boldsymbol{y}^\top\boldsymbol{H})^\top, \kappa_x\boldsymbol{Q}_x + \kappa_y\boldsymbol{H}^\top\boldsymbol{H} \right),\label{eq:xkk}\\
  \kappa_x|\boldsymbol{x}, \boldsymbol{y} &\sim \text{Gamma} \left( \alpha_x + \frac{m - 2}{2}, \beta_x + \frac{1}{2}\boldsymbol{x}^\top\boldsymbol{Q}_x\boldsymbol{x} \right),\label{eq:kxk}\\
  \kappa_y|\boldsymbol{x}, \boldsymbol{y} &\sim \text{Gamma} \left( \alpha_y + \frac{m}{2}, \beta_y + \frac{1}{2} \left(\boldsymbol{y} - \boldsymbol{H}\boldsymbol{x} \right)^\top \left(\boldsymbol{y} - \boldsymbol{H}\boldsymbol{x} \right) \right)\label{eq:kkx}.
\end{align}

\subsection{Scale-dependent features}
  In order find a decomposition of spatial data to recognize their dominant scale-dependent features, the data are smoothed on multiple scales, and the differences between consecutive smooths is calculated.
  Let $\boldsymbol{S}_{\lambda}$ be a roughness penalty smoother defined by $\boldsymbol{S}_{\lambda} = (\mathbf{I}_n + \lambda\boldsymbol{Q})^{-1}$, where $\lambda$ is the smoothing scale and $\boldsymbol{Q}$ is a generic spatial-weight matrix for the respective spatial data (according to equation~\eqref{eq:regularQ}).
  We smooth the spatial data $\boldsymbol{x}$  with $\boldsymbol{S}_{\lambda}\boldsymbol{x}$.
  Considering a sequence of smoothing scales, $0 = \lambda_1 < \lambda_2 < \ldots < \lambda_L = \infty$, such that $\boldsymbol{S}_{\lambda_1}\boldsymbol{x} = \boldsymbol{x}$ defines the identity mapping and $\boldsymbol{S}_{\lambda_L}\boldsymbol{x} = \boldsymbol{S}_{\lambda_{\infty}}\boldsymbol{x}$ the overall mean of $\boldsymbol{x}$.
  Then, $\boldsymbol{x}$ can be represented as differences of consecutive smooths: $\boldsymbol{x} = \boldsymbol{S}_{\lambda_1}\boldsymbol{x} - \boldsymbol{S}_{\lambda_2}\boldsymbol{x} + \boldsymbol{S}_{\lambda_2}\boldsymbol{x} - + \ldots - \boldsymbol{S}_{\lambda_{\infty}}\boldsymbol{x} + \boldsymbol{S}_{\lambda_{\infty}}\boldsymbol{x}$.
 Scale-dependent details are formalized as $\boldsymbol{z}_{\ell} = \left(\boldsymbol{S}_{\lambda_{\ell}} - \boldsymbol{S}_{\lambda_{\ell+1}} \right)\boldsymbol{x}$ for $\ell = 1, \ldots, L-1$ and $\boldsymbol{z}_L = \boldsymbol{S}_{\lambda_{\infty}}\boldsymbol{x}$.
  Exploiting the sparse structure of the spatial-weight matrix $\boldsymbol{Q}$, we solve the linear system $(\mathbf{I}_n + \lambda\boldsymbol{Q})^{-1}\boldsymbol{x}$ and decompose the spatial data as
\begin{equation}\label{eq:detaildecomp}
  \boldsymbol{x} = \sum_{\ell = 1}^{L-1} \left( \boldsymbol{S}_{\lambda_{\ell}}\boldsymbol{x} - \boldsymbol{S}_{\lambda_{\ell+1}}\boldsymbol{x} \right) + \boldsymbol{S}_{\lambda_{\infty}}\boldsymbol{x} = \sum_{\ell=1}^{L} \boldsymbol{z}_{\ell}.
\end{equation}

  A meaningful multiresolution decomposition requires a careful selection of smoothing scales.
  We follow~\citet{Pasanen2013}, who introduced the concept of the scale-derivative, defined by
\begin{equation}
  \frac{\partial \boldsymbol{S}_{\lambda}}{\partial \log\lambda} = \boldsymbol{D}_{\lambda}\boldsymbol{x}\nonumber.
\end{equation}
  The logarithmic scale for $\lambda$ is motivated by the fact that for increasing smoothing levels, the difference between successive values for $\lambda$ has to become wider to have an effect on the smooth.
  The derivative is
\begin{equation}\label{eq:scalederivative}
    \boldsymbol{D}_{\lambda} \boldsymbol{x} = \lim_{\lambda' \to \lambda} \frac{\boldsymbol{S}_{\lambda'}\boldsymbol{x} - \boldsymbol{S}_{\lambda}\boldsymbol{x}}{ \log{\lambda'} - \log{\lambda} } = \lambda (\mathbf{I}_n + \lambda \boldsymbol{Q})^{-1} \boldsymbol{Q} (\mathbf{I}_n + \lambda \boldsymbol{Q})^{-1}\boldsymbol{x}.
\end{equation}
  For efficient computation, this derivative can be reduced to solve two linear systems, with a single Cholesky factorization.
  Values for $\lambda_2, \ldots, \lambda_{L-1}$ can then be chosen as the local minima of $||\boldsymbol{D}_{\lambda}\boldsymbol{x}||$.
  \cite{Pasanen2013} use the Euclidean norm, which leads to smoothing scales representing averaged feature extents.
  As an alternative, we propose the maximum norm which seem to improve the detection of local and anisotropic features; see the illustration in Section~\ref{sec:compnorms} for a detailed comparison.

\subsection{Feature width-extents}\label{subsec:featsize}
  The application of multiresolution decomposition to spatial data recognizes their dominant scale-depen\-dent features.
  However, it gives us no estimate for the width of the extent of the scale-dependent features.
  To assess this width-extent, we calculate the empirical variogram and optimize corresponding Mat\'ern variogram function parameters.
  We estimate these parameters directly from the details, which are the differences between smooths.
  Subsequently, we can assess the width-extent of scale-dependent features with the spatial data-driven effective-range parameter.
  Therefore we assume the scale-dependent features to be a realization of a weakly stationary spatial process $\{ Z(\boldsymbol{s}): \boldsymbol{s} \in \mathcal{D} \subseteq \mathbb{R}^2 \}$, i.e.,
\begin{align}
  \text{E}(Z(\boldsymbol{s}_1)) &\equiv \mu\nonumber,\\
  \text{COV}(Z(\boldsymbol{s}_1), Z(\boldsymbol{s}_2)) &= cov(\boldsymbol{s}_1 - \boldsymbol{s}_2)\nonumber
\end{align}
  and thus $\text{Var}(Z(\boldsymbol{s}_1) - Z(\boldsymbol{s}_2)) = 2\gamma(\boldsymbol{s}_1 - \boldsymbol{s}_2)$, for all locations $\boldsymbol{s}_1, \boldsymbol{s}_2 \in \mathcal{D}$~\citep{Cressie93}.
  Thereby $cov$ denotes a covariance function, $2\gamma$ the variogram and $\gamma$ the semi-variogram function and it holds $\gamma(\boldsymbol{h})=cov(\boldsymbol{0})-cov(\boldsymbol{h})$.
  The difference $\boldsymbol{h} = \boldsymbol{s}_1 - \boldsymbol{s}_2$ expresses the spatial lag between two spatial locations.
  Often the variogram is parametrized with a nugget effect $\theta_3$: $\gamma(\boldsymbol{h}) \to \theta_3$ as $\boldsymbol{h} \to \boldsymbol{0}$ (if it exists); the partial sill $\theta_2$: the difference of the maximum value of the semi-variogram and the nugget effect; and the range $\theta_1$: parametrizing the distance where the variogram reaches its maximum (if it exists).
  In order to estimate the variogram parameters based on the spatial data, we calculate first the empirical variogram, e.g.~\citep{Matheron62},
\begin{equation}
  2\hat{\gamma}(\boldsymbol{h}) = \frac{1}{N_J} \sum_{(i, j) \in J } \left( Z(\boldsymbol{s}_i) - Z(\boldsymbol{s}_j) \right)^2\nonumber,
\end{equation}
  where $J = J(\boldsymbol{h}) = \{ (i, j): \boldsymbol{s}_i - \boldsymbol{s}_j \in T(\boldsymbol{h}) \}$, $N_J = \text{card}\{J\}$ and $T(\boldsymbol{h})$ is a specified tolerance region in $\mathbb{R}^2$~\citep{Cressie93, Deutsch98}.
  Based on the empirical variogram, we fit a variogram model, optimizing the respective parameters for a Mat\'ern variogram function.
  A Mat\'ern function is suitable for estimating the parameters of the individual smoothing based components, because of its parameterization with an additional smoothing parameter $\theta_4$.
  We can calculate the effective range $\theta_{\text{eff}}$, which is defined as the distance where the covariance function attains 5\% of the partial sill~\citep{Banerjee2003}.
  For the calculation of the empirical variogram and fitting the variogram function in \textsc{R} we use the packages \textbf{fields}~\citep{Nychka2020} and \textbf{gstat}~\citep{Pebesma2004}.

  \citet{Pasanen2018}~propose to assess the extent of the features for each scale-dependent component in the context of a roughness penalty, specifically a Nadaraya–Watson smoother using a Gaussian kernel, by matching the feature extents with the respective kernel widths.
  By the conceptual similarities of kernel smoothing and spatial process interpolation~\citep{Nychka2000} we expect similar results when matching kernel widths or when assessing ranges of covariance or variogram functions.
  However, variogram functions are much more flexible as we fit adaptable parametric families to the data.
  Furthermore, the variogram approach offers more versatility for data with anisotropic features~\citep{Journel1982}.
  In our experience, the Mat\'ern function is especially useful as it allows the estimation of the smoothing parameter, which increases the accuracy of the effective range.

  For non-normal spatial data, the recent work of~\citet{Oman2019} provides an alternative to the variogram estimation outlined above.
  In one-dimensional problems, such as time-series, the same approach can be used, but instead of variogram functions, the autocorrelation is estimated.

\subsection{Posteriori credibility analysis}
  We assign credibility to the dominant features and uncertainty to the smoothing scales and the width-extents.
  Uncertainty intervals can be constructed for the different smoothing scales by replacing $\boldsymbol{x}$ in equation~\eqref{eq:scalederivative} and applying it to each posterior sample from the model described in Fig.~\ref{fig:model}.
  With such uncertainty intervals, different scales can be identified.
  Thereby, we expect for small smoothing scales wide uncertainty intervals and for larger smoothing scales narrow intervals.
  As the posterior mean separates the assumed observational noise from the spatial data and is somewhat smoother than the posterior samples, we expect to detect more uncertainty for smoothing scales between zero and one on the $\text{log}_{10}$ scale.

  To assign probabilities to the details and their respective features, we use probability maps as in the scale-space analysis, for example, pointwise (PW) credibility maps, where every location $i=1,\dots,n$ of the $\ell$th detail $\boldsymbol{z}_{\ell}$ is allocated into one of three disjoint subsets according to the posterior probability of $\boldsymbol{z}_{\ell,i}$:
  $I^b=\{ i: \text{P}(\boldsymbol{z}_{\ell,i} > 0\mid \boldsymbol{y}) \ge \alpha \}$ or $I^r=\{ i: \text{P}(\boldsymbol{z}_{\ell,i} < 0\mid \boldsymbol{y}) \ge \alpha \}$ in which the components $\boldsymbol{z}_{\ell,i}$ pointwise differ credibly from zero or $I^g=\{1,\dots,n\}\setminus \left(I^b \cup I^r\right)$.
  As PW maps treat every location independently, these can sometimes exhibit only tiny islands of credibility.
  Contemporary approaches, on the other hand, assign credibility more conservatively to locations but maximize the connectedness of credible locations, as described by \citet{Erasto2005}.

  Similar to constructing uncertainty intervals for smoothing scales, we can construct uncertainty intervals for the effective-range parameters of the scale-dependent features.
  Therefore, we consider each calculated detail for each posterior sample of the spatial data and apply the procedure described in Section~\ref{subsec:featsize}.
  It returns the same number of estimated effective-range parameters corresponding to the number of posterior samples, enabling us to construct an uncertainty interval.

  \subsection{Computational and implementation aspects}\label{subsec:compdetails}
  For the efficient implementation of this spatial multiresolution method, we rely on the sparse structure of the precision and respective spatial-weight matrices.
  In particular, we use the compressed sparse row format of a matrix~\citep{Tewarson73, Bulucc2009}.
  To store a matrix in $\mathbb{R}^{n \times n}$ with $w$ non-zero elements we need therefore $w$ reals and $w+n+2$ integers compared with $n \times n$ reals.
  The main computational challenge in the outlined method is the calculation of the inverse of the spatial-weight matrix $\boldsymbol{Q}$ or, equivalently the smoothing matrix $\boldsymbol{S}_\lambda$.
  Here, for simplicity, we assume $\boldsymbol{Q}$ is a symmetric positive-definite matrix featuring the aforementioned sparse structure.
  Instead of intuitively solving $\boldsymbol{Q}\boldsymbol{x} = \boldsymbol{b}$, we proceed by factorizing $\boldsymbol{Q}$ into $\boldsymbol{U}^\top\boldsymbol{U}$, such that $\boldsymbol{U}$ is an upper triangular matrix with positive diagonal entries, known as the Cholesky factor of $\boldsymbol{Q}$.
  The linear systems $\boldsymbol{U}^\top\boldsymbol{y} = \boldsymbol{b}$ and $\boldsymbol{U}\boldsymbol{x} = \boldsymbol{y}$ are then solved by forward- and backward-solve respectively.
  To reduce so-called fill-in of the Cholesky factor $\boldsymbol{U}$, we permute the columns and rows of $\boldsymbol{Q}$ according to a permutation $\boldsymbol{P}$,~i.e., $\boldsymbol{T}^\top\boldsymbol{T} = \boldsymbol{P}^\top\boldsymbol{Q}\boldsymbol{P}$, where $\boldsymbol{T}$ is an upper triangular matrix.
  Further details on numerical methods for sparse matrices are outlined in \citet{Liu81, Duff86, Dongarra98} and for the statistical software \textsc{R} sparse matrices are described in \citet{Furrer2010}.

\section{Illustration}\label{sec:illustration}
\begin{figure}[t]
  \centerline{\includegraphics[width=\textwidth]{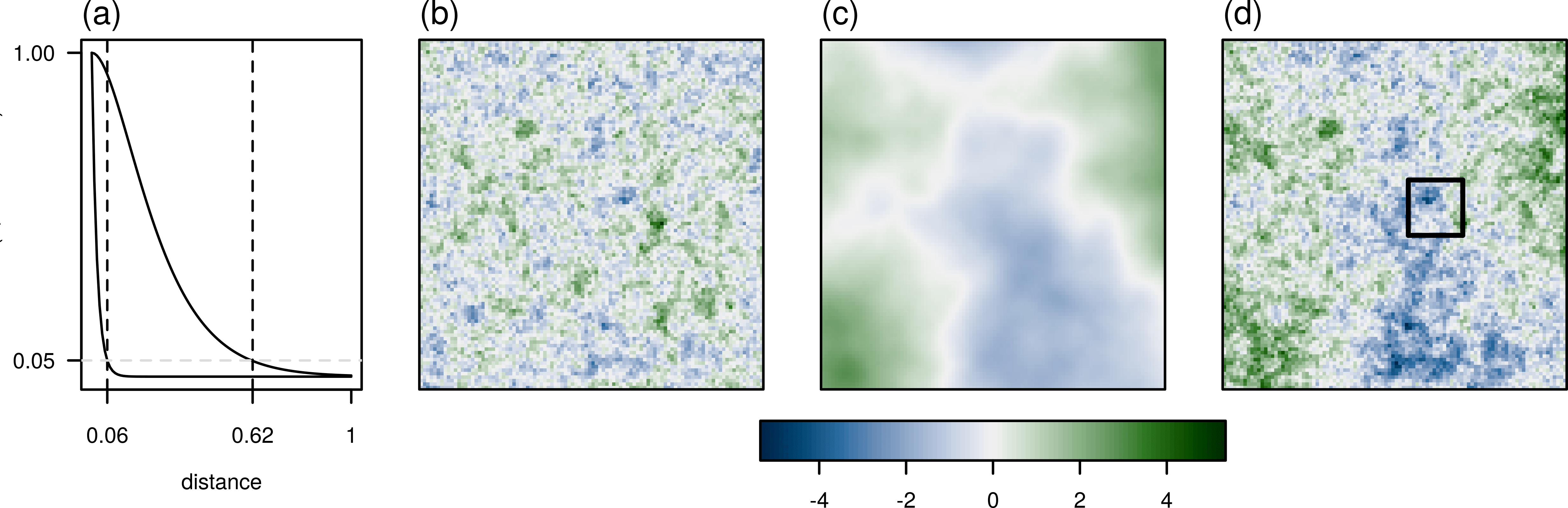}}
  \caption{Visualization of the first simulation setup.
           \textbf{(a)} Mat\'ern covariance functions used to simulate the data in~(b) and~(c), where the distance is with respect to the unit interval;
           \textbf{(b)} simulated spatial fields based on a Mat\'ern covariance function with $\theta_{\text{eff}} = 0.06$;
           \textbf{(c)} simulated spatial fields based on a Mat\'ern covariance function with $\theta_{\text{eff}} = 0.62$;
           \textbf{(d)} the sum of the simulated fields (b) and (c) and additional white noise; the area in the black rectangle is removed, and the respective grid locations are treated as missing values.}
  \label{fig:area_sampleoutna}
\end{figure}

  We illustrate the outlined method with simulated data on a regular grid of size $100 \times 100$ over $[0 ,1] \times [0 ,1]$.
  We sample two spatial fields based on Mat\'ern covariance functions with different parametrizations, i.e.,~with range $2$ and $12$, partial sill $1$ each, nugget $0$ each, and smoothness $0.5$ and $1.8$, respectively.
  The corresponding covariance functions have an effective range of 0.06 and 0.62, respectively (Fig.~\ref{fig:area_sampleoutna}~(a)).
  Imitating additive spatial data, we add together these two simulated components and additional white noise to one spatial field (Fig.~\ref{fig:area_sampleoutna}~(b)-(d)).
  The additional white noise represents observational noise, to simulate observed spatial data as closely as possible.
  We introduce missing values by removing an area of $15 \times 15$ grid points near the center of the simulated data.
  We apply each step of the \emph{feature identification} method to the data with missing values and the data without missing values to highlight the effect of these missing values on the scale-derivative, the dominant features and their width-extents.
  For a unique distinction between the analysis of the data with and without missing values, we denote in the following parameters, variables, and details for the data without missing values with an additional prime.

\begin{figure}
  \centerline{\includegraphics[width=\textwidth]{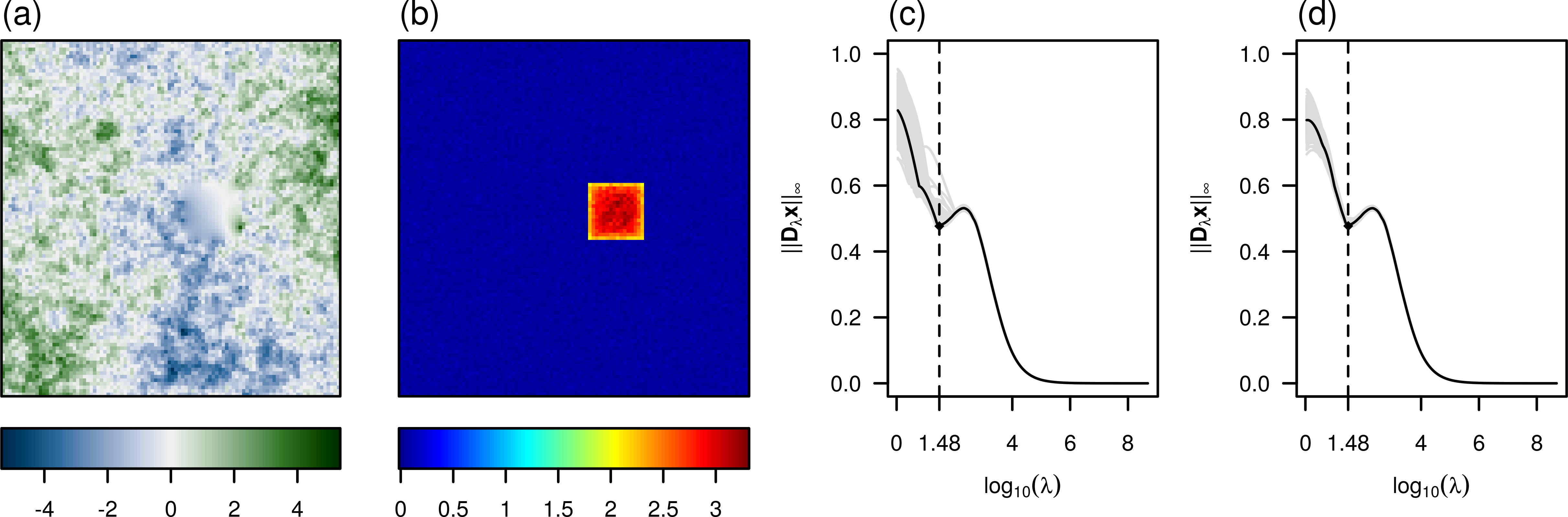}}
  \caption{Visualization of resampled data with missing values, the different level of uncertainty for locations with missing values and respective scale-derivatives.
           \textbf{(a)} The posterior mean $\text{E}(\boldsymbol{x}|\boldsymbol{y})$ based on resampled draws;
           \textbf{(b)} the width of 90\% posterior credibility interval of the posterior draws;
           \textbf{(c)}-\textbf{(d)} the scale-derivative maximum vector norms of the posterior mean  $\text{E}(\boldsymbol{x}|\boldsymbol{y})$ (black) and the respective scale-derivatives of the individual posterior samples (gray) for based on the simulated data with missing values (c), and for the full dataset (d).}
  \label{fig:resampled_sampleoutna}
\end{figure}

  We resample $\boldsymbol{x}$ (for the observed data with missing values) using the adjusted full-conditional distributions given in equations~\eqref{eq:xkk},~\eqref{eq:kxk} and~\eqref{eq:kkx}.
  Similarly, we resample $\boldsymbol{x}'$ (for the observed data without missing values) using the full-conditional distributions given in equations~\eqref{eq:normalgammacanonical},~\eqref{eq:gammaQ} and~\eqref{eq:gammaQ2}.
  The respective sampler is configured with the vaguely informative hyperparameters, $\alpha_x = 1$, $\alpha_y = 10$, $\beta_x = 0.1$ and $\beta_y = 1$, and after a burn-in phase of length 10'000, a chain of length 1'000 is considered.
  Standard tools such as trace and autocorrelation are taken into account to assure the convergence of the sample chain~\citep{Brooks98}.
  Figure~\ref{fig:resampled_sampleoutna}~(a) shows the posterior mean $\text{E}(\boldsymbol{x}|\boldsymbol{y})$ based on the samples, and~(b) the width of the 90\% posterior interval.
  Additionally, we compare the posterior distribution of $\kappa_y$ with the posterior distribution of $\kappa_{y'}$, and we find that $\kappa_{y'}$ has a marginally increased width.
  The grid-pointwise posterior mean smooths the posterior samples slightly.
  We make use of this smoothing by assuming that it accounts for observational noise in the spatial data.
  We can observe this by comparing the simulated field in Fig.~\ref{fig:area_sampleoutna}~(d) with the posterior mean in Fig.~\ref{fig:resampled_sampleoutna}~(a).
  For the locations with missing values, the width of the 90\% posterior credibility interval is substantially wider for the locations with missing values (Fig.~\ref{fig:resampled_sampleoutna}~(b)), as additional uncertainty is introduced by sampling the missing values.

  We calculate for a sequence of lambdas the scale-derivative using the posterior mean $\text{E}(\boldsymbol{x}|\boldsymbol{y})$ in equation~\eqref{eq:scalederivative}.
  We minimize the scale-derivative for the maximum vector norm that finds a local minimum for these simulated spatial data.
  In contrast, the Euclidean norm finds none for the same simulated data.
  These scale-derivatives, as well as the scale-derivatives based on the individual posterior samples, are visualized in Fig.~\ref{fig:resampled_sampleoutna}~(c) and for the complete data in panel~(d).
  Through this procedure, we obtain one scale $\lambda_2 = 30$ with 95\% uncertainty interval $(4, 45)$ and analogously for the simulated data without missing values $\lambda_2' = 30$ with $(3, 40)$.
  These smoothing scales are equal, where the uncertainty interval for the complete data is narrower as there is less uncertainty in the spatial data.
  For both cases, with and without missing values, $\lambda_1 = 0$ and $\lambda_3 = \lambda_{\infty} = \infty$ complete the set of smoothing scales.

\begin{figure}[H]
  \centerline{\includegraphics[width=\textwidth]{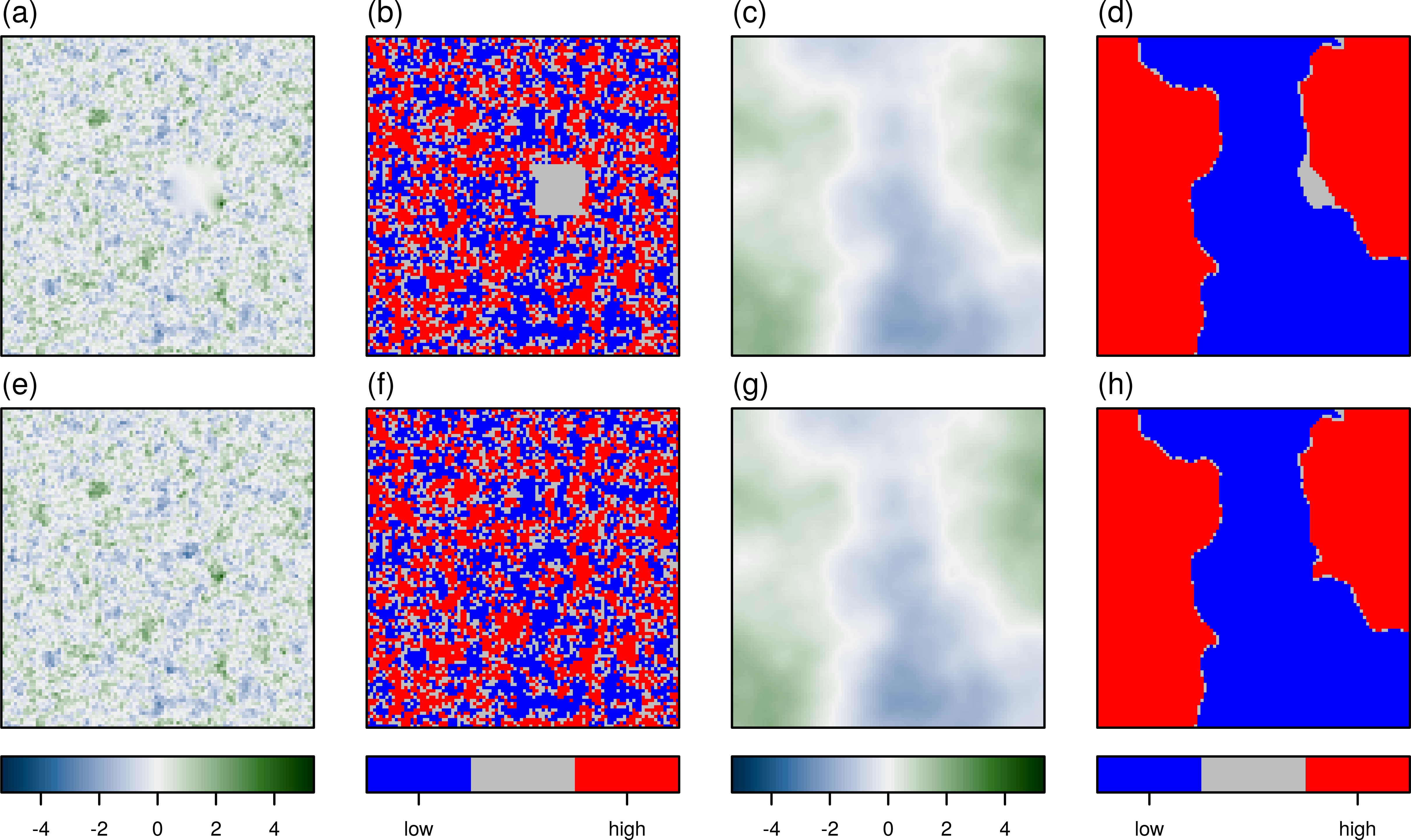}}
  \caption{Multiresolution details for spatial data with and without missing values.
           \textbf{(a)} $\text{E}(\boldsymbol{z}_1|\boldsymbol{y})$ summarized by the posterior sample mean;
           \textbf{(b)} PW maps of $\boldsymbol{z}_1$;
           \textbf{(c)} $\text{E}(\boldsymbol{z}_2|\boldsymbol{y})$ summarized by the posterior sample mean;
           \textbf{(d)} PW maps of $\boldsymbol{z}_2$;
           \textbf{(e)}-\textbf{(h)} are the analogous visualizations to~(a)-(d) for the spatial data without missing values.
           Each legend applies for both panels in the same column.}
  \label{fig:details_sample}
\end{figure}

  Based on these derived smoothing scales, we decompose the posterior samples of the simulated spatial data into two details according to equation~\eqref{eq:detaildecomp} with $\boldsymbol{z}_1 = \boldsymbol{S}_{\lambda_1}\boldsymbol{x} - \boldsymbol{S}_{\lambda_2}\boldsymbol{x}$ and $\boldsymbol{z}_2 = \boldsymbol{S}_{\lambda_2}\boldsymbol{x} - \boldsymbol{S}_{\lambda_3}\boldsymbol{x}$ (the overall mean $\boldsymbol{z}_3 = \boldsymbol{S}_{\lambda_{\infty}}\boldsymbol{x} = 0.03$ completes the decomposition).
  The resulting details and PW maps are shown in panels~(a)-(d) of Fig.~\ref{fig:details_sample} when controlling for sampling missing values; the respective details based on the data without missing values are shown in panels~(e)-(h).
  We can recognize dominant features in the two simulated spatial fields in the posterior means of the first and second detail and can assess a good separation into the two components.
  A comparison of the detail decomposition without missing values with the one with missing values (Fig.~\ref{fig:details_sample}) shows that this separation is similar for both cases.
  Moreover, for the second detail, the features in the locations with missing values are partially reconstructed, and the credibility analysis yields similar PW maps.

\begin{figure}[H]
  \centerline{\includegraphics[width=\textwidth]{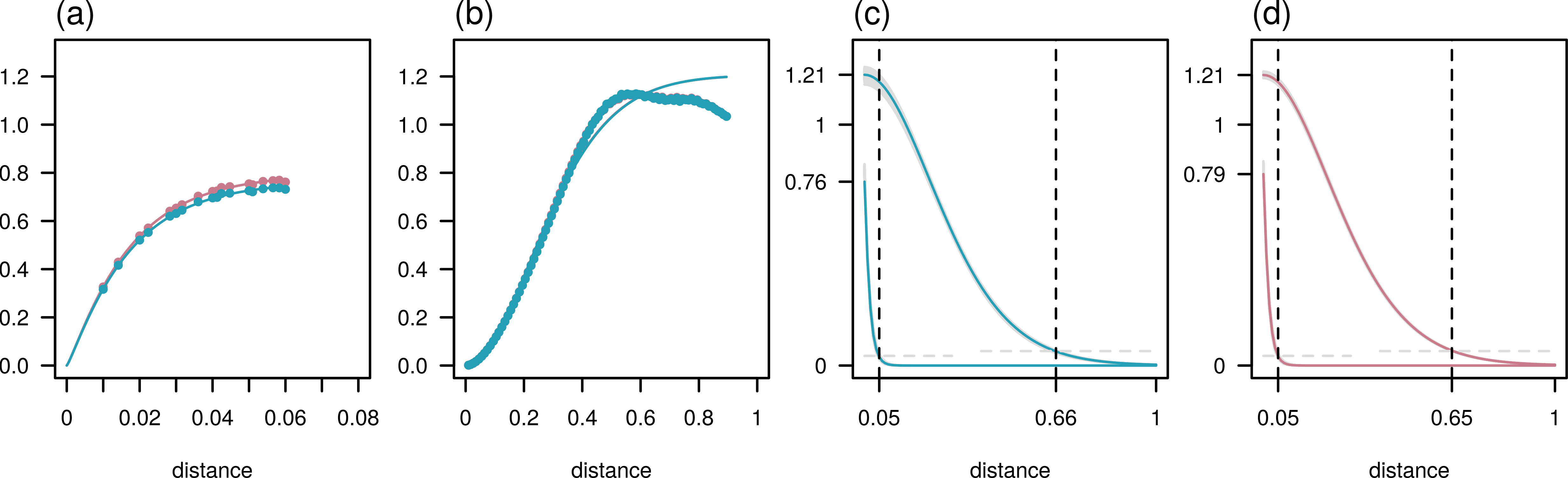}}
  \caption{Visualization of estimated variogram and fitted covariance functions.
    \textbf{(a)} and \textbf{(b)} the empirical variograms and fitted Mat\'ern variogram functions for the posterior mean of the details $\boldsymbol{z}_1$ and $\boldsymbol{z}_2$;
    \textbf{(c)} and \textbf{(d)} the covariance functions parametrized with the respective variogram estimates from panel \textbf{(a)} and \textbf{(b)} with the according effective range;
    \textbf{(c)} shows with vertical dashed lines for the details $\boldsymbol{z}_1$ and $\boldsymbol{z}_2$ the resulting effective ranges;
    \textbf{(d)} shows with vertical dashed lines for the details $\boldsymbol{z}'_1$ and $\boldsymbol{z}'_2$ the resulting effective ranges.
    Gray dashed horizontal lines indicate 5\% of the partial sill.
    In all panels, the color red indicates values based on the posterior means of the complete spatial data and blue those with missing values.
    Note, the distances in the individual panels are with respect to the unit interval.}
  \label{fig:variogram_sim}
\end{figure}

  Next, we estimate the effective range of the dominant features in the derived details $\text{E}(\boldsymbol{z}_1|\boldsymbol{y})$ and $\text{E}(\boldsymbol{z}_2|\boldsymbol{y})$, summarized by their posterior sample mean.
  We calculate the empirical variogram, fit a Mat\'ern variogram function to estimate its parameter, and calculate the effective range of the corresponding covariance function.
  For the respective summarized details $\text{E}(\boldsymbol{z}_1|\boldsymbol{y})$ and $\text{E}(\boldsymbol{z}_2|\boldsymbol{y})$, the variogram function fits are presented in Fig.~\ref{fig:variogram_sim}~(a) and~(b).
  We can construct uncertainty intervals for the range, partial sill and smoothness parameter of both details $\boldsymbol{z}_1$ and $\boldsymbol{z}_2$ based on the data with missing values as well as for the details $\boldsymbol{z}_1'$ and $\boldsymbol{z}_2'$ based on the complete data (Table~\ref{tab:illustraion}).
  In each detail, we estimate no nugget effect, and the partial sill therefore corresponds to the sill parameter.
  Furthermore, we set for the smoothness parameter an upper bound of value 5, as suggested in the \textbf{gstat} package.
  The smoothness parameter is best estimated based on the shortest distances.
  However, these are relatively wide on a regular equispaced grid.
  The numerical conditions to estimate the smoothness parameter are therefore not optimal.
  Nevertheless, including the estimate of this smoothness parameter increases the quality of the resulting effective ranges.
  In general, all the presented estimates are different with respect to their uncertainty intervals for detail $\boldsymbol{z}_1$ and $\boldsymbol{z}_2$.
  The respective uncertainty intervals based on the spatial data with missing values are, in all cases, wider than for the complete data.
  This also propagates into the resulting effective ranges.
  These effective ranges are close to the ones on which the simulated fields originate.
  With the sampling of the missing values, it is possible to obtain close results, but with slightly more uncertainty.
  The decomposition by differences of smooths proves useful as a filter of large-scale and small-scale effects.
  It is possible to recognize the dominant scale-dependent features and assess its respective width-extents, meaning that the dominant features are identified.

\begin{table}[ht]
\centering
\caption{Optimized Mat\'ern variogram function parameters and resulting effective-range for the different details summarized by their posterior sample means and the respective 95\% uncertainty intervals.
           For $\boldsymbol{z}_2$, the \textbf{gstat} optimizer set $\hat{\theta}_4$ to predefined maximum possible value.}
\label{tab:illustraion}
\begin{tabular}{r|ll|ll|ll|ll|l}
   & \multicolumn{2}{|l}{ range } & \multicolumn{2}{|l}{ (partial) sill } & \multicolumn{2}{|l}{ smoothness } & \multicolumn{2}{|l|}{ effective range } & true \\detail & $\hat{\theta}_1$ & 95\% CI & $\hat{\theta}_2$ & 95\% CI & $\hat{\theta}_4$ & 95\% CI & $\hat{\theta}_{\text{eff}}$ & 95\% CI & $\theta_{\text{eff}}$ \\
  \hline
\hline
$\boldsymbol{z}_1$ & 0.015 & (0.014, 0.018) & 0.76 & (0.79, 0.83) & 0.6 & (0.5, 0.6) & 0.05 & (0.05, 0.05) & \multirow{2}{*}{0.06} \\
  $\boldsymbol{z}_1'$ & 0.015 & (0.014, 0.018) & 0.79 & (0.8, 0.84) & 0.6 & (0.5, 0.6) & 0.05 & (0.05, 0.05) \\
  $\boldsymbol{z}_2$ & 0.081 & (0.079, 0.082) & 1.21 & (1.19, 1.23) & 5 & (-,-) & 0.66 & (0.65, 0.66) & \multirow{2}{*}{0.62}  \\
  $\boldsymbol{z}_2'$ & 0.08 & (0.08, 0.081) & 1.21 & (1.2, 1.21) & 5 & (-,-) & 0.65 & (0.65, 0.66) \\
   \hline
\end{tabular}
\end{table}

\subsection{Comparing scale selection using the maximum and Euclidean norm}\label{sec:compnorms}
  Although the maximum and Euclidean norm are equivalent in the sense of $||\boldsymbol{x}||_{\infty} \le ||\boldsymbol{x}||_2 \le \sqrt{n}||\boldsymbol{x}||_{\infty}$ for $\boldsymbol{x} \in \mathbb{R}^n$, both norms lead to different local minima of the scale-derivative $||\boldsymbol{D}_{\lambda}\boldsymbol{x}||$, defined in equation~\eqref{eq:scalederivative}.
  The minima based on the Euclidean norm represent global average feature extents, whereas the maximum norm emphasizes local features.
  To illustrate these differences, we look at two particular scenarios with a second and third simulation setup.

\begin{figure}[H]
    \centerline{\includegraphics[width=\textwidth]{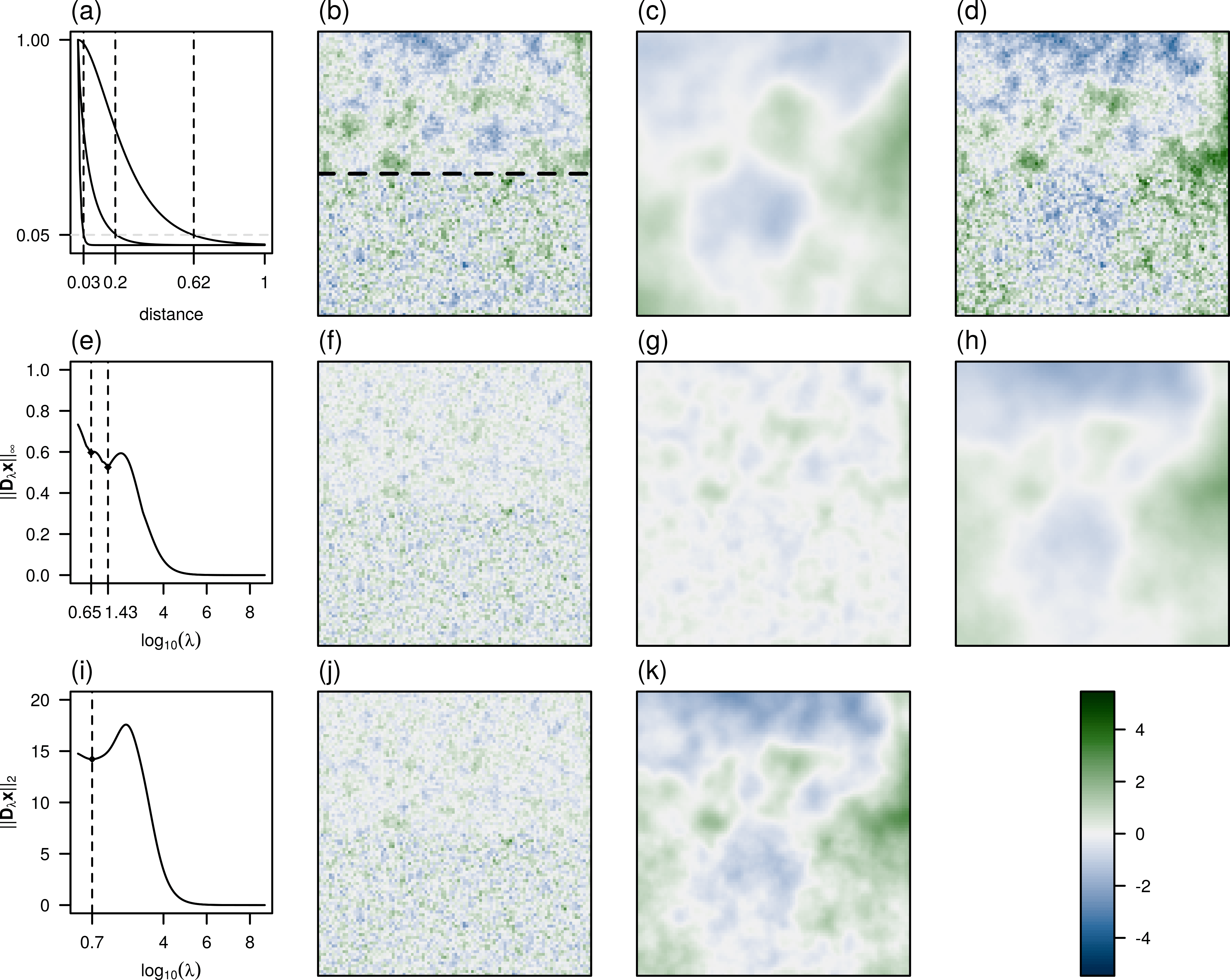}}
    \caption{Visualization of the second simulation setup, the scale-derivatives for the different norms and the respective multiresolution details.
             \textbf{(a)} Mat\'ern covariance functions used to simulate the data in~(b) and~(c), where the distance is with respect to the unit interval;
             \textbf{(b)} lower half of simulated spatial field based on a Mat\'ern covariance function with $\theta_{\text{eff}} = 0.03$ and upper half of simulated spatial field based on a Mat\'ern covariance function with $\theta_{\text{eff}} = 0.20$;
             \textbf{(c)} simulated spatial field based on a Mat\'ern covariance function with $\theta_{\text{eff}} = 0.62$;
             \textbf{(d)} the sum of the simulated fields (b) and (c) and additional white noise;
             \textbf{(e)} the scale-derivative maximum vector norms for the posterior mean $\text{E}(\boldsymbol{x}|\boldsymbol{y})$;
             \textbf{(f)}--\textbf{(h)} scale-dependent details summarized by their posterior sample means calculated with the scales based on the minima of the maximum norm, with effective-ranges $0.03$, $0.12$ and $0.61$;
             \textbf{(i)} the scale-derivative Euclidean vector norms for the posterior mean $\text{E}(\boldsymbol{x}|\boldsymbol{y})$;
             \textbf{(j)}--\textbf{(k)} scale-dependent details summarized by their posterior sample means calculated with the scales based on the minima of the Euclidean norm, with effective ranges $0.03$ and $0.56$.}
    \label{fig:comp_2}
  \end{figure}

  In the second simulation setup, we consider data where some dominant features only exist in a subarea to mimic processes that exhibit a local feature.
  We simulate such data by sampling from three Mat\'ern covariance functions with different parameters; we use range $1$, $8$ and $12$, partial sill $1$ each, nugget $0$ each, and smoothness $0.5$, $0.35$ and $1.8$, respectively.
  The corresponding effective-ranges are 0.03, 0.20 and 0.62 (Fig.~\ref{fig:comp_2}~(a)).
  To construct subareas with different dominant features, we use the first covariance parametrization to simulate the lower half of the area (below the dashed line in Fig.~\ref{fig:comp_2}~(b)),  and we use the second parametrization to simulate the upper part.
  The third sampled spatial field and additional white noise complete these example data (Fig.~\ref{fig:comp_2}~(c)-(d)).
  The maximum norm detects both scales and results in three scale-dependent details with different dominant features (Fig.~\ref{fig:comp_2}~(e)-(h)).
  Corresponding effective-range estimates are accurate for the first ($0.03$) and the last detail ($0.61$).
  The middle ($0.12$) is less accurate due to the remaining artifacts of the small scale features in the lower half of the data.
  The Euclidean norm averages the dominant features of the first two scales and only detects two components with different dominant features (Fig.~\ref{fig:comp_2}~(i)-(k)).
  The effective range estimate of the small detail ($0.03$) is driven by the true small scale and hence, accurate, whereas the effective-range estimate of the large scale detail is underestimated.

\begin{figure}[H]
    \centerline{\includegraphics[width=\textwidth]{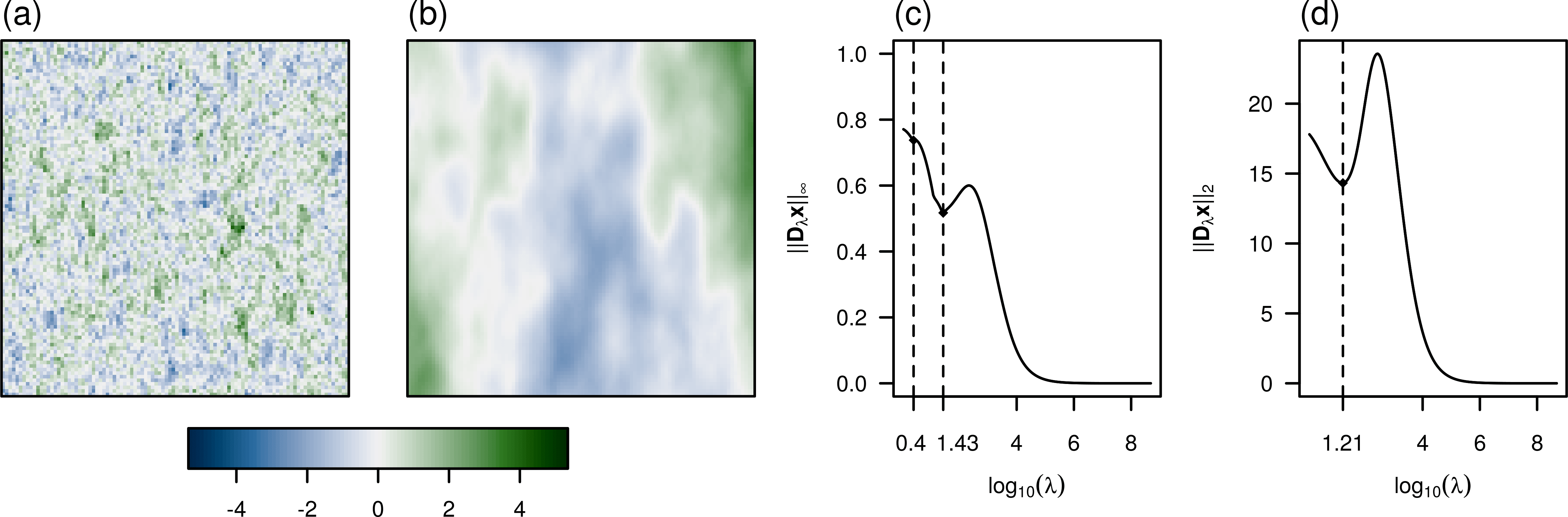}}
    \caption{Visualization of the third simulation setup, spatial data with anisotropy and scale-derivatives with respect to the maximum and the Euclidean norm.
             \textbf{(a)}--\textbf{(b)} simulated small and large scale anisotropic spatial field;
             \textbf{(c)}--\textbf{(d)} the scale-derivatives using the maximum norm and  the Euclidean norm for the posterior mean $\text{E}(\boldsymbol{x}|\boldsymbol{x})$.}
    \label{fig:comp_1}
  \end{figure}

  In the third simulation setup, we consider data from an anisotropic process leading to features with elliptical shapes.
  We simulate data with such features with the covariance functions shown in Fig.~\ref{fig:area_sampleoutna}.
  However, we scale the $x$-dimension by a factor of two for the simulation of the data.
  Hence, the resulting features are prolonged in the $y$ direction relative to the $x$ direction by a factor two (Fig.~\ref{fig:comp_1}~(a)--(b)).
  For this data, the maximum norm finds two scales (Fig.~\ref{fig:comp_1}~(c)), one representing the features longer and one its shorter axis.
  Conversely, the Euclidean norm finds a scale between those two scales (Fig.~\ref{fig:comp_1}~(d)), representing the average extents of the elliptical features.

\section{Identifying dominant features to find relevant interaction ranges of species}\label{sec:application}
  In this section, we describe an application of the outlined \emph{feature identification} method to an ecological problem.
  First, we present the observed spatial data.
  Then we motivate the ecological problem of capturing different sizes of areas where species communities interact and outline the calculation of biodiversity indices.
  Third, we describe the details of the data's pre-processing and focus on the application of the \emph{feature identification} method.
  Finally, we discuss and compare the results obtained here with those reported in the literature.

\subsection{Remote sensing data}
\begin{figure}
  \centerline{\includegraphics[width=\textwidth]{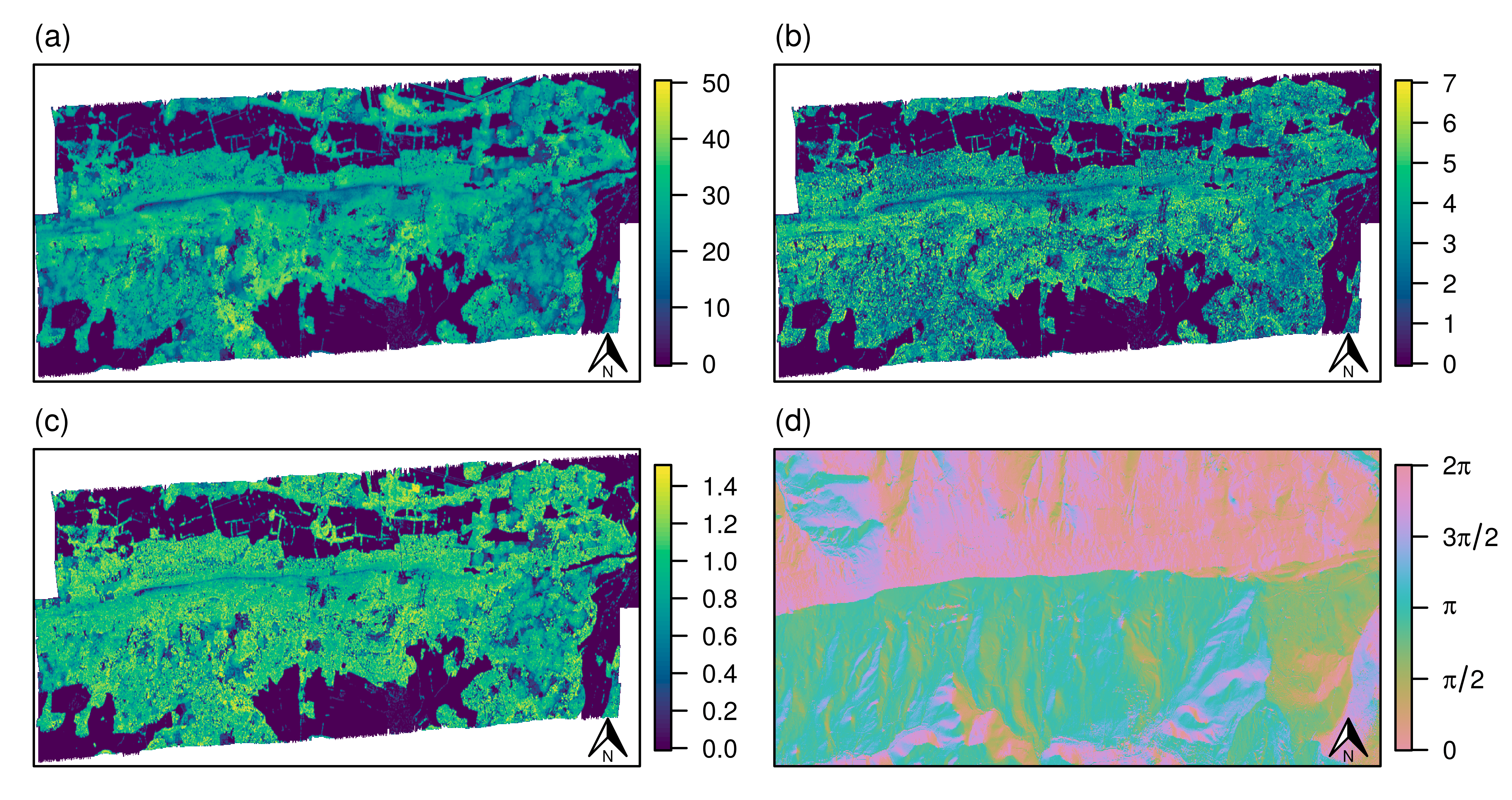}}
  \caption{Morphological traits of the mountain Laegeren.
           The area of interest has 400 grid points in N--S direction and 1'100 grid points in E--W direction.
           \textbf{(a)} Canopy height (CH) in~$m$;
           \textbf{(b)} plant area index (PAI) in~$m^2/m^2$;
           \textbf{(c)} foliage height diversity (FHD) index;
           \textbf{(d)} the topographic variable aspect of the mountain Laegeren in radians.}
  \label{fig:laegerentraits}
\end{figure}

  A team from the University Zurich, led by the head of the Remote Sensoring Laboratories, Prof.\ Dr.\ M.\ E.\ Schaepman, used light detection and ranging (LiDAR) scanning to detect the height, density, and form of trees in forests.
  They developed precise methods to systematically monitor variability in biodiversity indices based on remote-sensing data~\citep{Schneider2014, Morsdorf2009}.
  Such data were obtained from the Laegeren mountain site near Zurich, Switzerland.
  The investigated area is about 2 $\times$ 6 km in extent, and the natural vegetation of the Laegeren mountain is a lightly managed beech-dominated forest, which has a relatively high diversity of tree species, age, and diameter~\citep{Eugster2007}.
  The raw LiDAR measurements were combined to describe specific structural characteristics of the Laegeren mountain forest using morphological traits.
  For a detailed description and explanation of the selected traits, we refer to~\citet{Schneider2014, Schneider2017}.
  The morphological traits we consider are canopy height (CH) of trees in meters~($m$), plant area index (PAI) in square meter per square meter~($m^2/m^2$) and foliage height diversity (FHD), a measurement of the amount of foliage at various levels above the ground, converted into an index.
  Figure~\ref{fig:laegerentraits}~(a)-(c) visualizes these three morphological traits.
  Apparent features in the traits are the mountain ridge parallel to the $x$-axis (E--W) and an area of juvenile trees in the south-western part, which is the aftermath of a storm disturbance~\citep{Schneider2017}.
  These traits are calculated per pixel of an overlying equispaced grid, consisting of 400 $\times$ 1'100 grid points at distances of six meters.
  With such continuous area-based data, contributions of more than one individual or species to the trait in a single-pixel are possible.
  Therefore, a pixel does not represent a direct link to an individual specimen.
  Figure~\ref{fig:laegerentraits} also shows that the traits do not entirely cover the overlying grid of the rectangular area.

\subsection{Ecological context}
  In plant ecology, it is assumed that vegetation is spatially heterogeneous and features emerge as a result of interaction between species.
  Thereby interaction happens on different scales between different communities of species~\citep{Greig79}.
  In the morphological traits of the mountain Laegeren forest, a community of species could be defined by single trees, by small troops of trees which expanded together or larger assemblages of trees which define ecosystems of different sizes.
  The interaction range of species within such a community is therefore of particular interest, as these define dominant spatial features and thereby influence measures for its ecosystem functioning and performance~\citep{Liu2018}.

  Functional biodiversity indices are measures that aim to explain the susceptibility of the forest to climate change or its ecosystem functioning and performance.
  Various diversity definitions exist, as well as several indices to quantify those.
  We use the multidimensional functional diversity indices richness, divergence, and evenness~\citep{Villeger2008}.
  These multidimensional functional diversity indices are calculated based on multiple trait data and a moving-window approach (see~\ref{appendix:indices}.1 for details).
  The moving window maps neighboring pixels in a predefined radius, which has the advantage that intraspecific diversity is included and that it is independent of vegetation units, species, or plant functional types~\citep{Schneider2017}.
  When calculating these multidimensional functional diversity indices, the choice of the moving window radii is of paramount importance.
  To choose suitable radii, \cite{Schneider2017} proposed to calculate the indices for a reasonable range of radii and select the radius for which the overall average index curve has the steepest gradient.
  This procedure is typically applied to the entire study area, as well as to some subareas chosen based on expert knowledge.
  However, dependent on the size and resolution of the data, calculation of the indices takes a considerable amount of time, due to its computational efforts ($\mathcal{O}(n^2\log(C))$, where $C$ is the number of pixels inside a moving window, representing the size of a community per radius and different indices,~\cite{Chazelle1993, Pettie2002}).
  The computational burden of this approach lies in the construction of the indices, which is repeated for every considered radius with corresponding moving-window size.
  From an ecological perspective, it is interesting to understand, on which radii different features for these multidimensional functional diversity indices become manifest.

\subsection{Application}
  Motivated by such ecological questions, we show the feasibility of the \emph{feature identification} procedure outlined in Section~\ref{sec:mresd}, with the data and diversity indices introduced.
  Therefore, we apply this procedure to each morphological trait (CH, PAI, FHD) separately by first using the multiresolution decomposition to recognize their dominant scale-dependent features and secondly to determine the effective range of each scale-dependent feature.
  Assuming that the spatial dependency assesses relevant interaction between species, we interpret the effective range here as the interaction range, which defines the typical diameter of a community of species.
  As a result, we obtain the different interaction ranges of species for each morphological trait.
  Since these traits were observed from the same area and describe the forest's morphological properties, we can expect similar results per trait.
  The interaction range of each component is based on the dominant scale-dependent feature and we further propose that the interaction ranges are suitable candidates for the moving-window radius to calculate the multidimensional functional diversity indices.

  From an ecological perspective, it is also interesting to understand which environmental variables affect the variability of the resulting functional diversity.
  As previous studies of this area have shown, the variance of biodiversity indices based on morphological traits can be partially explained with topographic variables such as altitude, slope, and aspect~\citep{Schneider2017}.
  To take these factors into account, we detrend each morphological trait before the decomposition, according to these variables for the mountain Laegeren.
  The source for altitude, slope, and aspect (Fig.~\ref{fig:laegerentraits}~(d)) of the Laegeren mountain is https://geodata4edu.ethz.ch for the year 2018.
  To ensure the same resolution for these variables as for the morphological traits, we apply ordinary kriging for spatial interpolation~\citep{Cressie93}.
  Then, effective ranges can be derived based on the decomposition of these topographic variables and compared with the effective respectively the interaction ranges based on the morphological traits.
  Thereby, it is possible to put into context the derived effective ranges and subsequent radii and the variables from which these origins.

  In the following, we describe the individual steps to derive interaction ranges and moving-window radii for the diversity indices.
  The standardized residuals of the detrended morphological traits ensure equal weighting in the diversity indices and fulfill the theoretical assumptions of the multiresolution decomposition.
  Because of the eminent ridge in the forest area, the model depicted in Fig.~\ref{fig:model} with an IGMRF precision matrix of order one is well suited to represent the dependencies between locations.
  The size of the overlying regular grid of the Laegeren area determines the initial dimension of the precision matrix.
  As the traits do not entirely cover its overlying grid, we modify precision matrix, by eliminating corresponding columns and rows of $\boldsymbol{Q}$, see Section~\ref{sec:reggrid}.
  To reconstruct the traits, we calibrate the model with hyperparameters $\alpha_x = 0.1$, $\beta_x = 0.0005$ and $\alpha_y = 10$, $\beta_y = 1$, such that the prior distributions are vaguely informative.
  After a burn-in phase of 10'000 draws, we consider 1'000 samples for further analysis.

\begin{figure}
  \centerline{\includegraphics[width=\textwidth]{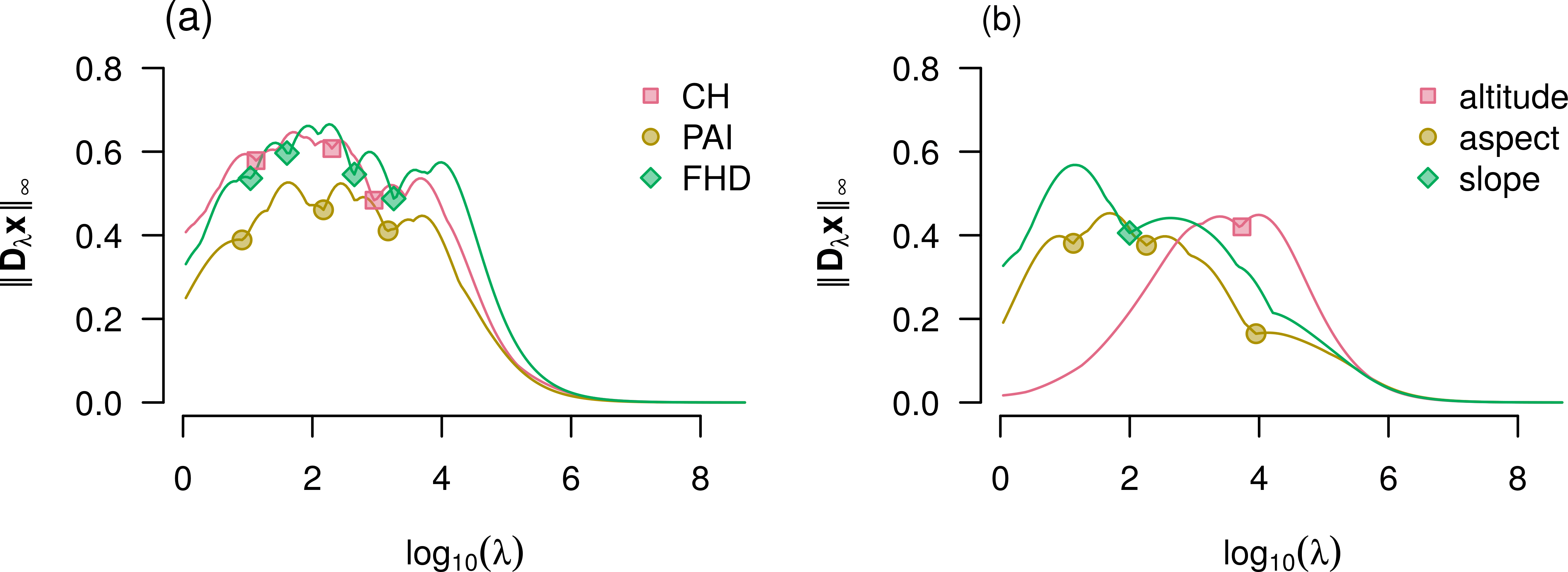}}
  \caption{Scale-derivatives for different lambdas, with squares, circles, and triangles showing the minima of the maximum norm for the individual datasets.
           \textbf{(a)} scale-derivatives for the morphological traits;
           \textbf{(b)} scale-derivatives for the topographic variables of the mountain Laegeren.}
  \label{fig:findscalesi}
\end{figure}

  We calculate the scale-derivatives (according to equation~\eqref{eq:scalederivative}) and determine local minima regarding the Euclidean and maximum vector norm for smoothing-scale candidates.
  This is done for the three morphological traits as well as for the three topographic variables and summarized in Table~\ref{tab:scales}.
  To determine smoothing scales using the scale-derivatives, we choose the local minima of the maximum norm (Fig.~\ref{fig:findscalesi}).
  Note that as the scales' size increases, the local minima are with respect to a wider range of scales.  Since the resulting difference of smooths between two scales becomes smaller as the scales become larger, for more details see~\cite{Pasanen2013}.
  Using the maximum norm is suitable for this data, as we can assume that the north-south slope, as well as the mountain ridge, introduce pronounced anisotropic features.
  As illustrated in Section~\ref{sec:compnorms}, the maximum norm detects the different dominant feature extents in such data; which is in accordance with the results in Table~\ref{tab:scales}.

\begin{figure}
  \centerline{\includegraphics[width=\textwidth]{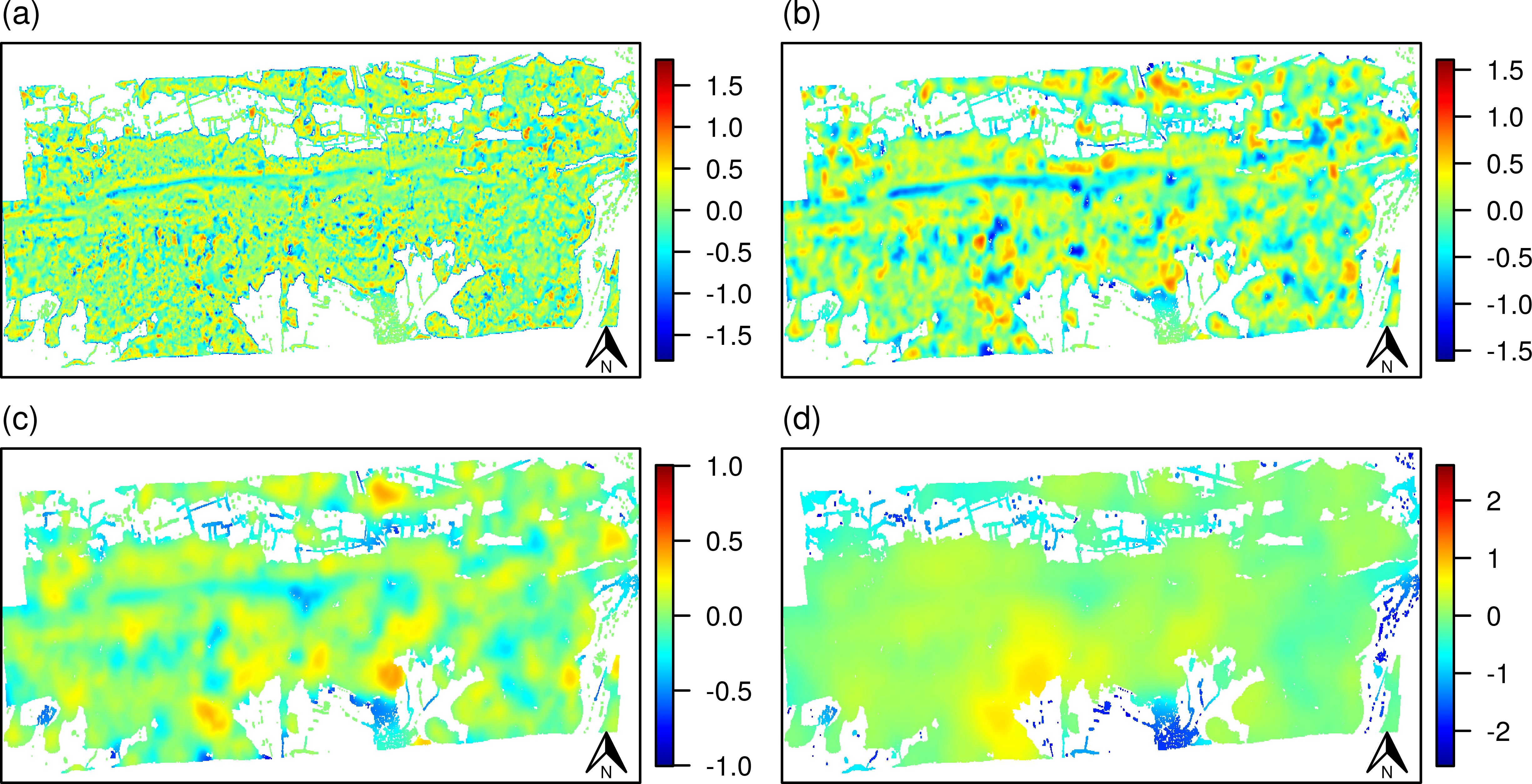}}
  \caption{Scale-dependent details summarized by their posterior sample means for the morphological trait CH.
           \textbf{(a)} $\text{E}(\boldsymbol{z}_1|\boldsymbol{y})$ for $\lambda_1 = 0,   \lambda_2 = 13$;
           \textbf{(b)} $\text{E}(\boldsymbol{z}_2|\boldsymbol{y})$ for $\lambda_2 = 13,  \lambda_3 = 200$;
           \textbf{(c)} $\text{E}(\boldsymbol{z}_3|\boldsymbol{y})$ for $\lambda_3 = 200, \lambda_4 = 897$;
           \textbf{(d)} $\text{E}(\boldsymbol{z}_4|\boldsymbol{y})$ for $\lambda_4 = 897, \lambda_{\infty}$.}
  \label{fig:ch_detail}
\end{figure}

\begin{figure}
  \centerline{\includegraphics[width=\textwidth]{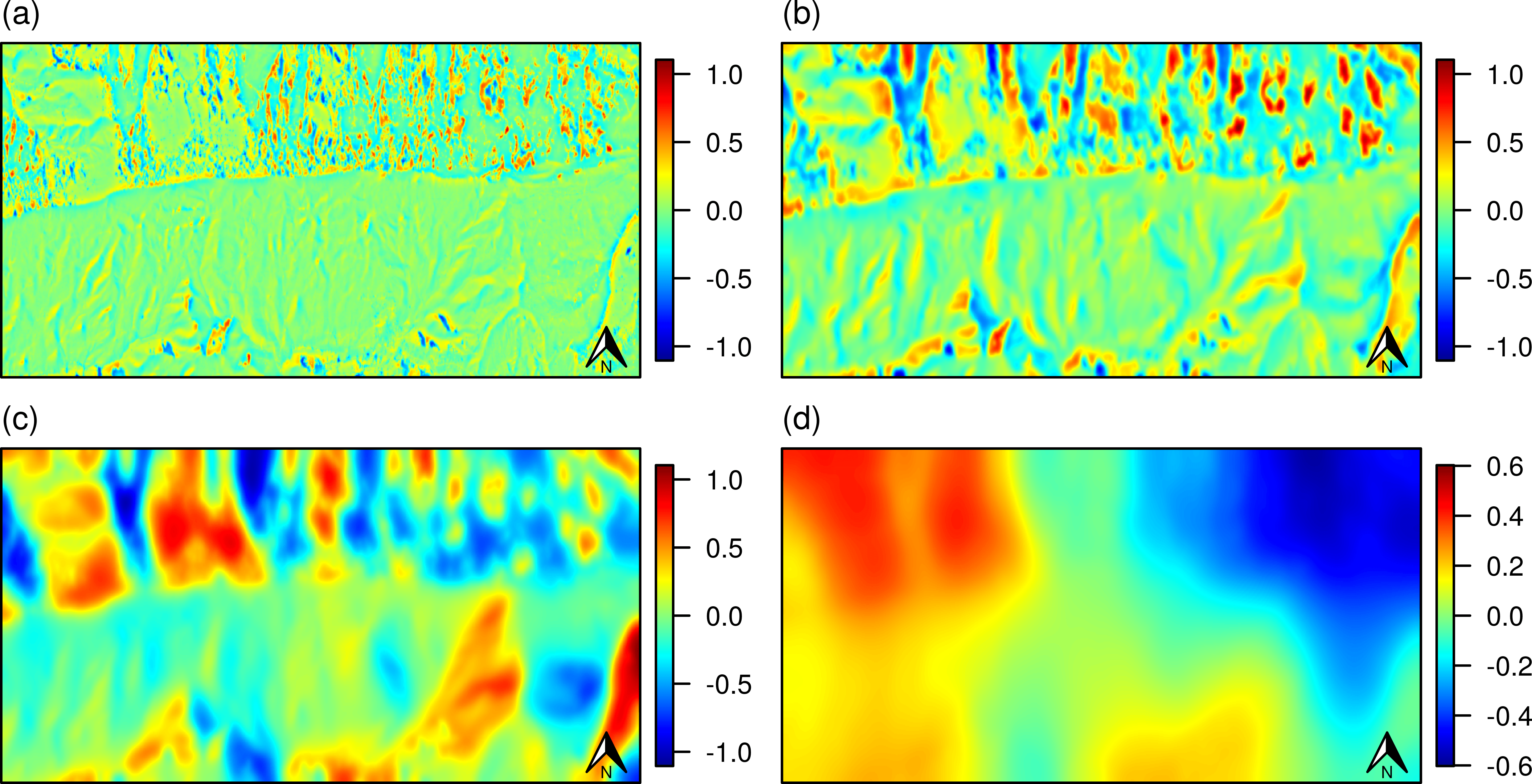}}
  \caption{Scale-dependent details summarized by their posterior sample means for the topographic variable aspect.
           \textbf{(a)} $\text{E}(\boldsymbol{z}_1|\boldsymbol{y})$ for $\lambda_1 = 0,   \lambda_2 = 13$;
           \textbf{(b)} $\text{E}(\boldsymbol{z}_2|\boldsymbol{y})$ for $\lambda_2 = 13,  \lambda_3 = 181$;
           \textbf{(c)} $\text{E}(\boldsymbol{z}_3|\boldsymbol{y})$ for $\lambda_3 = 181, \lambda_4 = 8955$;
           \textbf{(d)} $\text{E}(\boldsymbol{z}_4|\boldsymbol{y})$ for $\lambda_4 = 8955, \lambda_{\infty}$.}
  \label{fig:aspect_detail}
\end{figure}

  Next, we calculate the details of the multiresolution decomposition as differences of consecutive smooths, using the selected smoothing scales, see Figs~\ref{fig:ch_detail} and~\ref{fig:aspect_detail} for the visualization of the details for CH and aspect.
  Based on the details of all morphological traits and topographic variables, we calculate the empirical variograms and fit a Mat\'ern variogram function to estimate its respective effective range parameter, in number of pixels.
  To account for the anisotropic nature of the respective features, we use directional variograms in east--west (E--W) and north--south (N--S) direction.
  These empirical variograms and their parameters are estimated for the details based on the posterior mean as well as for the individual posterior samples, to calculate uncertainty intervals for the estimates.
  As we estimate the variogram parameters for each trait, each detail, and each posterior sample, we subsample the corresponding spatial data for computational competitiveness.
  To select a subsample, we can exploit the E--W and N--S directions of the dominant features given by the specific topology of the mountain Laegeren area.
  We estimate variogram functions in E--W direction based on a subsample of entire E--W-transects and analogous for the N--S directional variograms.
  To reliably and efficiently estimate the directional variogram function parameters, it is sufficient, to randomly select one of every three consecutive transects.
  This transect subsampling enables us to reasonably efficiently calculate the uncertainty intervals for each estimated effective-range parameter for a scale-dependent feature and select the different radius among all variables.

\begin{table}[]
\centering
\caption{Minima by scale-derivatives for Euclidean and maximum vector norms.}\label{tab:scales}
\begin{tabular}{l|llllll}
  $\text{norm}\backslash \text{trait/}\atop \text{variable}$ & CH      & PAI          & FHD               & altitude & aspect        & slope   \\[1mm] \hline\hline
  $||\boldsymbol{D}_{\lambda}\boldsymbol{x}||_2$        & 28           & 27           & 23                & 4800     & 12            & 1, 4800 \\
  $||\boldsymbol{D}_{\lambda}\boldsymbol{x}||_{\infty}$ & 13, 200, 897 & 8, 148, 1480 & 11, 40, 445, 1808 & 4917     & 13, 181, 8955 & 99      \\ \hline
\end{tabular}
\end{table}

\subsection{Results}
  A summary of the effective range, based on the estimated range, partial sill and smoothness parameter of a Mat\'ern covariance function and the respective 95\%-uncertainty intervals of the directional variogram functions is given in Table~\ref{tab:range}.
  The morphological trait details are each more distinctive in E--W direction than in N--S except for the most extensive details, including global trends of the area.
  This is in contrast to the topographic variables, for which no clear pattern emerge.
  CH and PAI are decomposed in four details, for the first two and the fourth CH has slightly more extensive,~i.e., larger or more expanded, effective-ranges and PAI for the third one.
  The morphological trait FHD shows an additional scale-dependent dominant feature compared with the other two traits, where CH-$\boldsymbol{z}_2$ and PAI-$\boldsymbol{z}_2$ are identified as a pronounced E--W elliptic feature and FHD-$\boldsymbol{z}_2$ is rather homogeneous but with smaller width-extent.
  Moreover, FHD-$\boldsymbol{z}_3$ and FHD-$\boldsymbol{z}_4$ are similar to $\boldsymbol{z}_2$ and $\boldsymbol{z}_3$ of CH and PAI, but also broader in extent.
  Also here, the identified global features are for all three morphological traits similar and most extensive in extent for PAI.
  Based on these trait details we identify five ``overall'' different width-extents to reduce the total number: 7.57 (CH-$\boldsymbol{z}_1$), 14.95 (FHD-$\boldsymbol{z}_2$), 34.21 (FHD-$\boldsymbol{z}_3$), 67.46 (PAI-$\boldsymbol{z}_3$), and 130.37 (FHD-$\boldsymbol{z}_4$), in number of grid points.

  In addition, we can estimate effective ranges based on the topographic variables; altitude and slope are decomposed in two and aspect in four details.
  The magnitude of the smoothing scale for the variable altitude implies one detail is describing regional and one global feature.
  We therefore consider an additional effective range 270.73 based on the altitude.
  The topographic variable aspect shows four details were aspect-$\boldsymbol{z}_1$ is similar to CH-$\boldsymbol{z}_1$, aspect-$\boldsymbol{z}_2$ is similar to FHD-$\boldsymbol{z}_3$, and aspect-$\boldsymbol{z}_3$ is between PAI-$\boldsymbol{z}_3$ and FHD-$\boldsymbol{z}_4$.
  These similarities show how the topographic variable aspect partially explains the respective dominant features, which is consistent with the analysis from~\cite{Schneider2017}.
  The topographic variable slope is decomposed into two details, in a local one, which is similar in width-extent to FHD-$\boldsymbol{z}_2$, and a global one; both are more pronounced in E--W direction.

  The 95\%-uncertainty intervals are non-overlapping for all range parameters within each spatial dataset.
  The uncertainty intervals of these estimates are, as expected, narrow for small effective ranges and broader for larger effective ranges.

  Finally, we can choose the moving-window radii for the functional diversity indices based on half of the estimated feature width-extents, multiplied by the grid point distance of six meters.
  We obtain the (rounded) set of radii $\{ 24\,m, 42\,m, 102\,m, 204\,m, 390\,m, 810\,m \}$.
  Thereof we can exclude the largest one, as its resulting indices would not be reasonable.
  To validate the quality of the derived effective ranges or the radii of for the functional diversity indices based on the \emph{feature identification}, we compare these with the expert-driven set of radii $\{ 12\,m, 60\,m, 240\,m \}$ from~\citet{Schneider2017}.
  We can assess that the \emph{feature identification} found a finer set of different radii.
  However, the method missed the smallest interaction range, which is between individual species.
  This is due to the pixel-based and not species-based data structure and the difficulties of estimating variogram function parameters for small features on regularly gridded data.

  The multidimensional indices are finally calculated based on these derived radii shown in~\ref{appendix:indices}.2 (Figs~\ref{fig:richness},~\ref{fig:evenness} and~\ref{fig:divergence}).
  In addition to the determination of moving-window radii, it is possible to link each radius to a specific variable, which can help to understand small- or large-scale effects from input data.

\begin{table}[!h]
\centering
\caption{Estimated effective range parameters of the respective morphological traits and topographic variables, in number of grid points.}
\label{tab:range}
\begin{tabular}{lll|rc}
  trait/variable & detail & direction & $\hat{\theta}_{\text{eff}}$ & 95\% CI \\
  \hline
\hline
CH & $\boldsymbol{z}_1$ & N--S &    6.37 & (5.90, 6.10) \\
   &  & E--W &    7.57 & (7.20, 7.40) \\
   \cline{2-5} & $\boldsymbol{z}_2$ & N--S &   19.36 & (19.00, 19.20) \\
   &  & E--W &   23.54 & (23.10, 23.70) \\
   \cline{2-5} & $\boldsymbol{z}_3$ & N--S &   38.64 & (38.30, 38.80) \\
   &  & E--W &   59.05 & (58.31, 60.51) \\
   \cline{2-5} & $\boldsymbol{z}_4$ & N--S & ${>}$400 & (${>}$400, ${>}$400) \\
   &  & E--W &  390.62 & (381.33, 397.74) \\
   \hline
PAI & $\boldsymbol{z}_1$ & N--S &    5.47 & (3.80, 4.00) \\
   &  & E--W &    6.16 & (4.30, 4.40) \\
   \cline{2-5} & $\boldsymbol{z}_2$ & N--S &   16.11 & (15.30, 15.60) \\
   &  & E--W &   19.74 & (19.10, 19.90) \\
   \cline{2-5} & $\boldsymbol{z}_3$ & N--S &   44.10 & (42.40, 45.80) \\
   &  & E--W &   67.46 & (66.00, 72.51) \\
   \cline{2-5} & $\boldsymbol{z}_4$ & N--S & ${>}$400 & (${>}$400, ${>}$400) \\
   &  & E--W &  382.71 & (357.22, 390.87) \\
   \hline
FHD & $\boldsymbol{z}_1$ & N--S &    5.79 & (4.70, 4.80) \\
   &  & E--W &    6.50 & (5.40, 5.60) \\
   \cline{2-5} & $\boldsymbol{z}_2$ & N--S &   13.44 & (12.90, 13.10) \\
   &  & E--W &   14.95 & (14.30, 14.60) \\
   \cline{2-5} & $\boldsymbol{z}_3$ & N--S &   25.01 & (24.60, 25.00) \\
   &  & E--W &   34.21 & (33.30, 34.60) \\
   \cline{2-5} & $\boldsymbol{z}_4$ & N--S &   61.41 & (57.51, 65.51) \\
   &  & E--W &  130.37 & (118.91, 132.51) \\
   \cline{2-5} & $\boldsymbol{z}_5$ & N--S & ${>}$400 & (${>}$400, ${>}$400) \\
   &  & E--W &  486.30 & (382.93, 503.07) \\
   \hline
altitude & $\boldsymbol{z}_1$ & N--S &  153.35 & (153.11, 153.41) \\
   &  & E--W &  270.73 & (268.42, 272.52) \\
   \cline{2-5} & $\boldsymbol{z}_2$ & N--S &  333.63 & (338.93, 339.13) \\
   &  & E--W &  489.98 & (509.15, 512.55) \\
   \hline
aspect & $\boldsymbol{z}_1$ & N--S &   10.04 & (6.20, 6.40) \\
   &  & E--W &    7.99 & (5.20, 5.40) \\
   \cline{2-5} & $\boldsymbol{z}_2$ & N--S &   26.66 & (25.60, 26.00) \\
   &  & E--W &   23.74 & (23.00, 23.30) \\
   \cline{2-5} & $\boldsymbol{z}_3$ & N--S &  119.15 & (116.61, 121.51) \\
   &  & E--W &   82.81 & (82.11, 83.71) \\
   \cline{2-5} & $\boldsymbol{z}_4$ & N--S & ${>}$400 & (${>}$400, ${>}$400) \\
   &  & E--W & ${>}$1'100 & (${>}$1'100, ${>}$1'100) \\
   \hline
slope & $\boldsymbol{z}_1$ & N--S &   11.18 & (10.90, 11.10) \\
   &  & E--W &   21.53 & (21.20, 21.70) \\
   \cline{2-5} & $\boldsymbol{z}_2$ & N--S &  142.63 & (204.72, 206.72) \\
   &  & E--W &  388.34 & (572.35, 592.66) \\
   \hline
\end{tabular}
\end{table}

\section{Discussion}\label{sec:discussion}
  The outlined procedure to identify features in spatial data has several interesting aspects.
  We show that our multiresolution decomposition is suitable for spatial data.
  The Bayesian framework of the decomposition permits resampling of missing values, with which it is possible to reconstruct scale-dependent spatial features with according credibility or uncertainty.
  On the other hand, by excluding areas of no interest, it is possible to precisely model and subsequently decompose the spatial data.
  The decomposition and recognition of the scale-dependent features make it possible to identify the variogram or covariance parameters of the underlying scale-dependent spatial process.
  The width-extent assessment for scale-dependent features with effective ranges is thereby more precise for larger details, based on larger smoothing scales; based on small ones, the smoothing parameter is difficult to estimate on regularly gridded data.

  The development of this method and its methodological choices favor structures and techniques that are computationally efficient.
  Most noteworthy in this context are the sparse structure of the precision matrix, the underlying hierarchical Bayesian model, as well as the transect subsampling scheme for directional variogram functions.
  Nonetheless, there is more potential computational gain, for example by using so-called tapering functions to estimate the effective range parameter for scale-dependent features with a maximum-likelihood approach of covariance functions.

  The application to scientifically relevant data enables new advances in finding different interaction ranges between species and defining communities of species.
  Notably, it was possible to respect the strong anisotropic characteristics of the dominant scale-dependent features, given by the topography of the area of interest.
  A final, careful ecological interpretation or explanation of these different communities would adequately describe the ecosystem functioning and therefore the scale-dependent features in the spatial data.

\section*{Acknowledgment}
  This work was supported by the Swiss National Science Foundation (SNSF 175529, P400P2\_186680, P2ZHP2\_174828) and by the University of Zurich Research Priority Program on Global Change and Biodiversity.
  We thank the reviewers for constructive comments on the manuscript.

\section*{Declarations of interest}
none

\section*{Author statement}
  Roman Flury: Conceptualization, Methodology, Data Curation, Visualization, Software, Writing - Original Draft preparation.
  Florian Gerber: Methodology (sampling missing values), Writing - Review \& Editing.
  Bernhard Schmid: Resources (Laegeren mountain data), Writing - Review \& Editing on ecological application.
  Reinhard Furrer: Supervision, Writing - Review \& Editing.

\bibliography{featsize_biblio.bib}

\newpage
\appendix

\section{Source Files}
\label{appen:code}
  Supplementary material is available in the git repository at:\\
  \url{https://git.math.uzh.ch/roflur/spatialfeatureidentification}.
  It contains the following source files.
\begin{itemize}
  \item README.txt: description of how to install devel packages, to access the data and run the illustration and application of this paper.
  \item source/: containing sources of devel packages that are used for running the code in analysis/.
  \item data/: containing \textsc{R}-scripts to load and transform input data
  \item analysis/: containing \textsc{R}-scrips to run illustration and application.
\end{itemize}

\section{Multidimensional functional diversity indices}\label{appendix:indices}

\subsection{Description}
  Functional richness (FRich) measures the extent of functional space which is occupied by a community of species.
  It is calculated as the convex hull volume in the space spanned by the multiple traits within the given radius of neighboring pixels~\cite{Villeger2008}.
  The definition implies an increase of the indices value with the radius of the window and is expected to follow on average a log curve for increasing radii~\citep{Schneider2017}.

  Functional divergence (FDiv) shows how species are distributed within the volume of the multidimensional functional trait space occupied by a community of, e.g., trees.
  To calculate multidimensional functional divergence, let $C$ be the number of pixels in a community, $dG_i$ the Euclidean distance between the $i$th pixel and the center of gravity and $\overline{dG}$ the mean distance of the $C$ pixels and the center of gravity.
  Then,
\begin{align}
  \Delta |d| &=  \frac{1}{C}\sum_{i=1}^{C} |dG_i - \overline{dG}|\nonumber,\\
  \text{FDiv} &= \frac{\overline{dG}}{\Delta|d| + \overline{dG}}\nonumber,
\end{align}
  such that all pixels are equally weighted (no abundances).

  Functional evenness (FEve) is calculated based on the minimum spanning tree (MST) of the $C$ pixels within a community.
  $\text{EW}_l$ denotes the (Euclidean) length of the $l$th branch of the MST, which is used to define the partially weighted evenness
\begin{equation}
  \text{PEW}_l = \frac{\text{EW}_l}{\sum_{l=1}^{C-1} \text{EW}_l}\nonumber
\end{equation}
  and the multidimensional functional evenness
\begin{equation}
  \text{FEve} = \frac{\sum_{l=1}^{C-1} \min{\Bigl( \text{PEW}_l, \frac{1}{C-1} \Bigr(} - \frac{1}{C-1} }{ 1 - \frac{1}{C-1} }.\nonumber
\end{equation}
  High functional richness for small window radii shows high diversity within communities and for large window radii high diversity between communities.

\subsection{Results}

  This section contains the visualization of the resulting multidimensional functional-biodiversity indices from the application in Section~\ref{sec:application}.

\begin{figure}
  \centerline{\includegraphics[width=\textwidth]{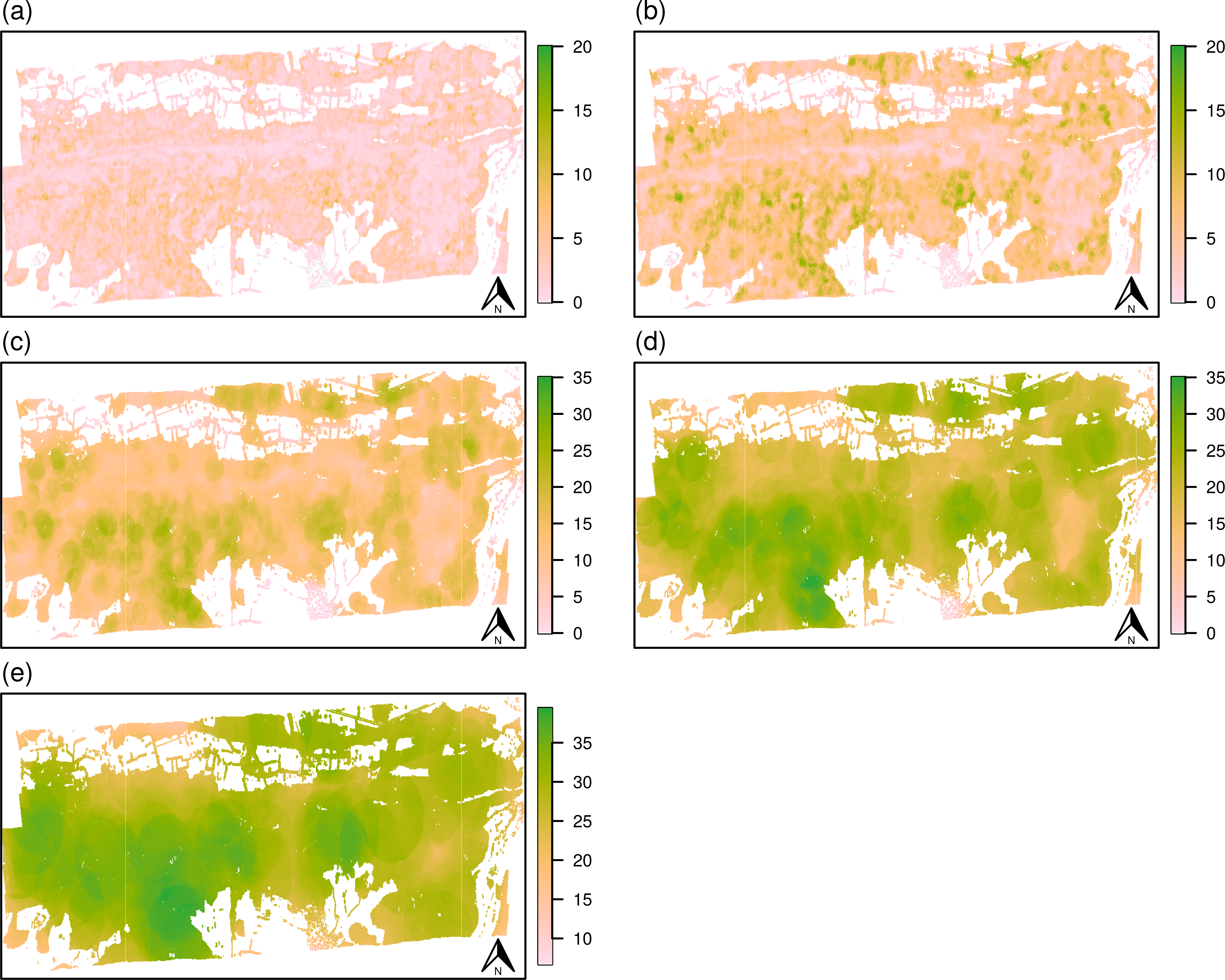}}
  \caption{Functional richness with moving-window radii $\{ 24\,m, 42\,m, 102\,m, 204\,m, 390\,m \}$.}
  \label{fig:richness}
\end{figure}

\begin{figure}
  \centerline{\includegraphics[width=\textwidth]{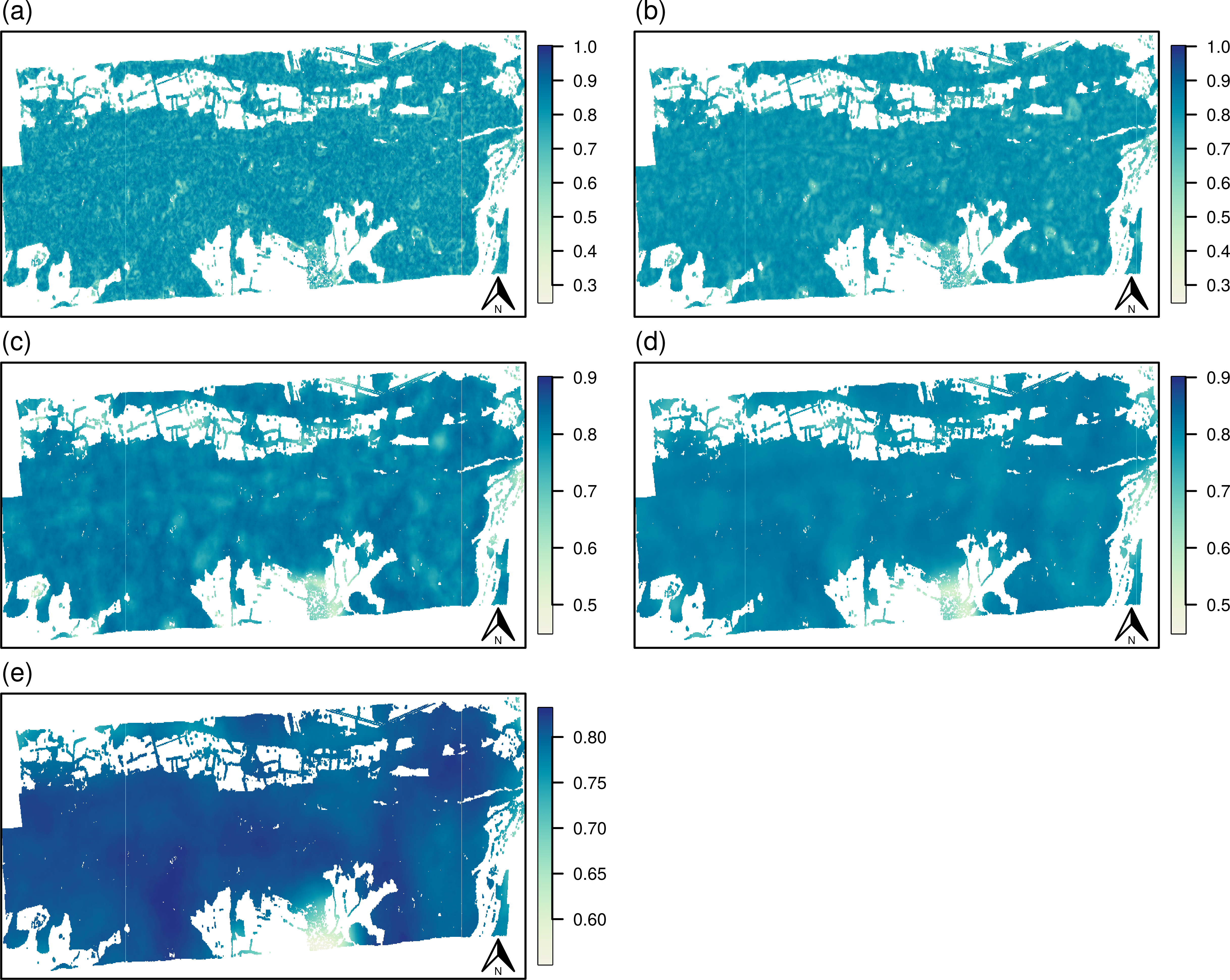}}
  \caption{Functional evenness with moving-window radii $\{ 24\,m, 42\,m, 102\,m, 204\,m, 390\,m \}$.}
  \label{fig:evenness}
\end{figure}

\begin{figure}
  \centerline{\includegraphics[width=\textwidth]{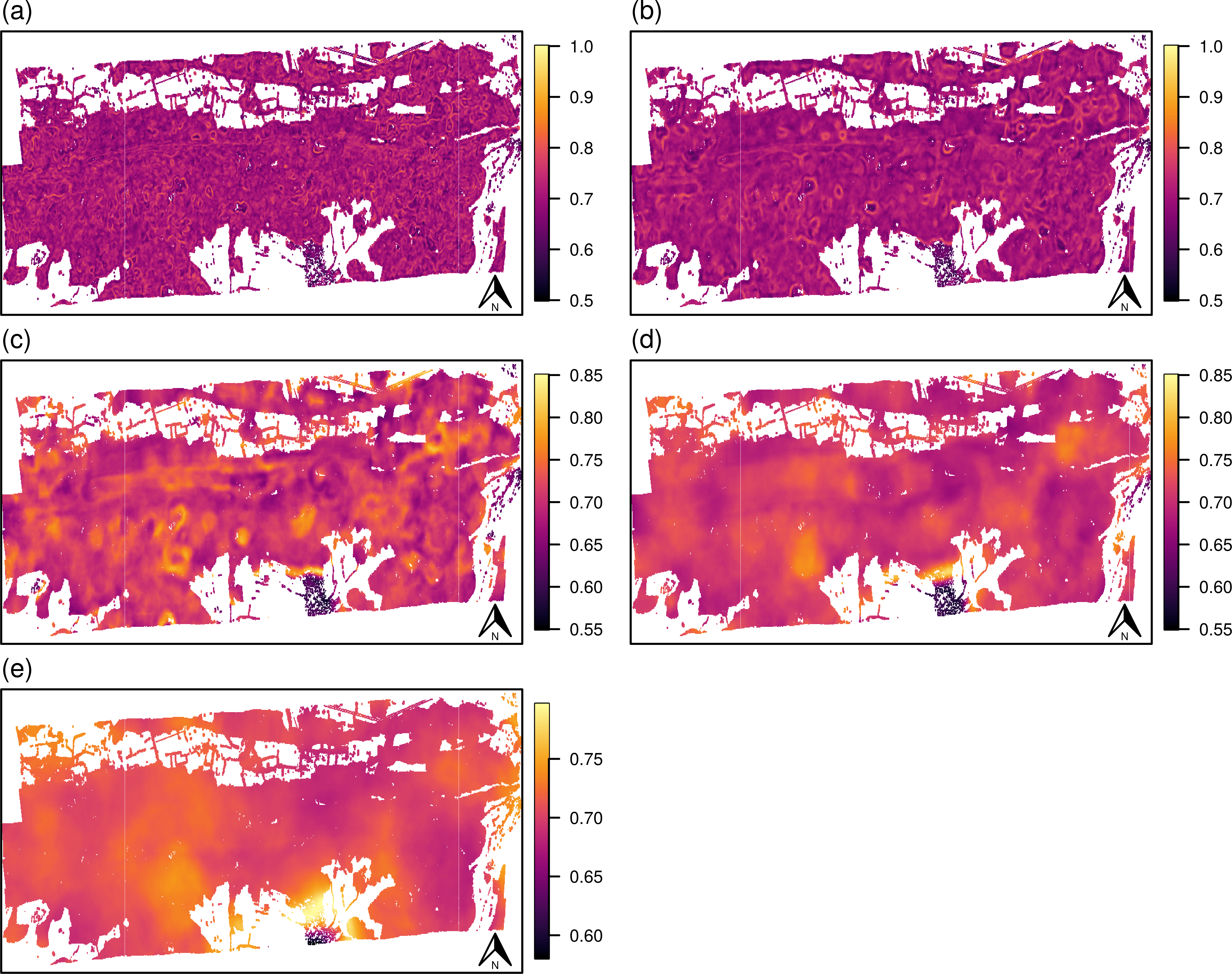}}
  \caption{Functional divergence with moving-window radii $\{ 24\,m, 42\,m, 102\,m, 204\,m, 390\,m \}$.}
  \label{fig:divergence}
\end{figure}

\end{document}